\documentclass[twocolumn]{aastex631}

\usepackage[version=4]{mhchem}

\usepackage{savesym}
\savesymbol{tablenum}
\usepackage{siunitx}
\restoresymbol{SIX}{tablenum}

\shorttitle{I. Observations and Radiative Transfer}
\shortauthors{James et al.}
\graphicspath{{./}{figures/}}
\begin{document}

\title{Revealing the Physical Conditions around Sgr A* using Bayesian Inference - I. Observations and Radiative Transfer}

\correspondingauthor{Tomas A. James}
\email{tjames@star.ucl.ac.uk}

\author[0000-0002-4240-4359]{Tomas A. James}
\affiliation{Department of Physics and Astronomy, 
University College London, 
Gower Street, London 
WC1E 6BT, UK } 

\author[0000-0001-8504-8844]{Serena Viti}
\affiliation{Leiden Observatory, 
Leiden University, 
PO Box 9513, NL-2300RA, 
Leiden, The Netherlands} 
\affiliation{Department of Physics and Astronomy, 
University College London, 
Gower Street, London 
WC1E 6BT, UK } 

\author{Farhad Yusef-Zadeh}
\affiliation{Department of Physics and Astronomy and CIERA, 
Northwestern University, 
Evanston, 
IL 60208, USA}

\author{Marc Royster}
\affiliation{Department of Physics and Astronomy and CIERA, 
Northwestern University, 
Evanston, 
IL 60208, USA}

\author[0000-0002-1737-0871]{Mark Wardle}
\affiliation{Department of Physics and Astronomy and Research Centre for Astronomy, 
Astrophysics and Astrophotonics, 
Macquarie University, 
Sydney, 
NSW 2109, Australia}



\begin{abstract}
We report sub-arcsecond ALMA observations between 272 - 375 \si{\giga\hertz} towards Sgr A*'s Circumnuclear disk (CND). Our data comprises 8 individual pointings, with significant \ce{SiO (8_{7} - 7_{6})} and \ce{SO (7 - 6)} emission detected towards 98 positions within these pointings. Additionally, we identify \ce{H_{2}CS (9_{1,9} - 8_{1,8})}, \ce{OCS (25 - 24)} and \ce{CH_{3}OH (2_{1,1} - 2_{0,2})} towards a smaller subset of positions. By using the observed peak line flux density together with a Bayesian Inference technique informed by radiative transfer models, we systematically recover the physical gas conditions towards each of these positions. We estimate that the bulk of the surveyed gas has temperature $\mathrm{T_{kin}} < 500$ \si{\kelvin} and density $\mathrm{n}_{\mathrm{H}} \lessapprox 10^{6}$ \si{\per\centi\meter\cubed}, consistent with previous studies of similar positions as traced by \ce{HCN} clumps. However, we identify an uncharacteristically hot ($\mathrm{T_{kin}} \approx 600$ \si{\kelvin}) and dense ($\mathrm{n}_{\mathrm{H}} \approx 10^{6}$ \si{\per\centi\meter\cubed}) source in the Northeastern Arm. This position is found to be approximately consistent with a gravitationally bound region dominated by turbulence. We also identify a nearby cold ($\mathrm{T_{kin}} \approx 60$ \si{\kelvin}) and extremely dense ($\mathrm{n}_{\mathrm{H}} \approx 10^{7}$ \si{\per\centi\meter\cubed}) position that is again potentially bound and dominated by turbulence. We also determine that the total gas mass contained within the CND is $M_{\mathrm{CND}} \approx 4 \times 10^{4}$ $M_{\odot}$. Furthermore, we qualitatively note that the observed chemical enrichment across large scales within the CND is consistent with bulk grain processing, though multiple desorption mechanisms are plausibly responsible. Further chemical modelling is required to identify the physical origin of the grain-processing, as well as the localised \ce{H_{2}CS} and \ce{OCS} emission.

\end{abstract}

\keywords{astrochemistry, galaxies: ISM, ISM: clouds, ISM: kinematics and dynamics, methods: observational, methods: statistical}

% \NewPageAfterKeywords

\section{Introduction} \label{sec:intro}

The Galactic Center, and within it the Central Molecular Zone, has long been identified as a uniquely extreme environment within the Milky Way owing to the $(4.154 \pm 0.014) \times 10^{6}$ M$_{\odot}$ supermassive black hole known as Sgr A* that resides there \citep{sgrABlackHole,sgrADistance}. Such an environment has important consequences for theories of galactic evolution and dynamics, as the Galactic Center can be used as a high-resolution laboratory for studying the nuclei of external galaxies. However, it also has a profound impact on theories of both ISM physics and chemistry under extreme conditions, especially when considering dense molecular gas as precursors to star formation.

Morphologically, the inner few \si{pc} of the Galactic Center is chaotic and multifaceted. Centrally, Sgr A* is situated within a trifecta of ionised gas streams known as the Galactic Center Mini Spiral, or GCMS \citep{gcms,gcms2}. Surrounding the GCMS is an extensive torus or ring like structure of turbulent molecular gas known as the Circumnuclear ring or Circumnuclear disk, herein referred to as the CND \citep[e.g.][]{Christopher2005}. The diffuse region between Sgr A* and the CND - known as the Central Cavity - is home to a collection of young and evolved stellar objects, including $\approx 100$ OB stars \citep{centralClusterOBStars,sgrACentralClusterIonisation,OBstarsDisk,sgrACentralCluster} that are thought to considerably increase the photoionisation rate within and around Sgr A* \citep{gcmsDarkMass,sgrACentralClusterIonisation}. 

Studies of the CND are complicated by a degree of uncertainty surrounding its exact morphology and kinematics. The extent of the CND is easily identified in low $J$-transitions of molecules such as \ce{CS}, \ce{H_{2}CO} and \ce{SiO} \citep{sgrAChemistry,tsuboiCNR} as well as dust continuum emission \citep{Lau_2013} and high $J$-transitions of molecules like \ce{^{12}CO} \citep{sgrACRIRLL}. Such observations identify the CND as extending radially from $\sim$1.6 pc outwards to $\sim$5--7 pc from Sgr A* \citep{Smith2014TheSignatures}, rotating at $\sim$110 \si{\kilo\meter\per\second} between 2--5 pc \citep{ghezStellarOrbits,Genzel2010}. 

However, the CND is not continuous or indeed uniform as occasional regions devoid of emission known as `gaps' are observed, in turn indicating that different molecules are mapping different regions of the CND. Notable amongst these gaps is that seen in \ce{HCN (1-0)} towards the North of Sgr A*. \citet{sgrANorthernGap} propose that this gap is caused by an infalling arm of the Galactic Center Mini Spiral pushing through the CND as it falls towards Sgr A* proper, producing \ce{OH} maser emission \citep{sgrAOHMaser} in the process. It is therefore clear that the gas conditions within the CND are highly non-uniform and position dependent.

In addition, the interior of the CND is highly irregular and inhomogeneous, exhibiting clump like structures of size $0.14 - 0.43$ \si{pc} distributed throughout when viewed in the near-IR \ce{HCN (1-0)} transition \citep{sgrACNDClumpHCN,sgrACNDClumpNeut,Christopher2005}. A number of theories regarding the physical conditions within these clumps exist, and \citet{Genzel2010} extensively summarises the two prevailing theories as the `transient' and `virial' scenarios. In summary, the transient scenario suggests that the clumps are typically warm ($> 100$ \si{K}), less dense ($< 10^{6}$ \si{\per\centi\meter\cubed}) and tidally unstable with transient lifetimes $\approx 10^{5}$ \si{yrs}. Conversely, the virial scenario states that these clumps are cooler ($< 100$ \si{K}), dense ($10^{7} - 10^{8}$ \si{\per\centi\meter\cubed}) and tidally stable with lifetimes $> 10^{7}$ \si{yrs}. Crucially, this stability affords these clumps the opportunity to begin in-situ star formation. The tidal instability, and therefore shorter lifetime estimate in the transient scenario, would not permit any gravitationally bound object formation. 

Within the detected clumps, \citet{tsuboiCNR} estimate that some of the relatively low density regions have gas densities in the range $0.9 - 6\times10^{4}$ \si{\per\centi\meter\cubed}. \citet{sgrAHCN} estimate denser \ce{HCN} bright cores to have densities in the range $0.1 - 2\times10^{6}$ \si{\per\centi\meter\cubed}. \citet{sgrAHCN}, and independently \citet{sgrAHCN2}, find that the \ce{HCN} clumps have conditions consistent with that outlined in the transient scenario. Crucially, such studies have not explored kinetic gas temperatures $> 250$ \si{\kelvin} and thus could fail to highlight hot and dense embedded sources. However, \citet{sgrAHCN} theorise that the CND is subject to such large tidal shear that local star formation is unlikely to occur within the CND in the absence of some triggering event. This tidal shear condition could be overcome should self-gravity dominate according to $\sim 10^{7}(r/\mathrm{pc})^{-3}$ \si{\per\centi\meter\cubed} \citep{sgra_bp_flows}. Assuming that the bulk of the CND is found between 1.5 -- 2 pc from Sgr A* as traced by \ce{HCN} \citep{Christopher2005}, then exceeding the tidal shear condition would require gas density between $ \approx 2 \times 10^{6}$ \si{\per\centi\meter\cubed} at $\mathrm{r} \approx 1.5$ pc, dropping to $ \approx 2 \times 10^{5}$ \si{\per\centi\meter\cubed} at $\mathrm{r} \approx 2$ pc. Satisfying this criterion is essential for in-situ star formation to occur within these clumps. 

C-type shocks originating from clump-clump collisions have been proposed as an efficient heating mechanism within the CND owing to the high observed temperatures of $\mathrm{T_{kin}} > 200$ \si{\kelvin} \citep{warmCNDGas, chemicalFeaturesCND}. However, star formation as a process also presents multiple opportunities to drive C-type shocks in to the surrounding gas, in turn chemically enriching the surrounding ISM in a similar manner to clump-clump collisions. \ce{SiO} and \ce{SO} - as well as other \ce{S}-bearing species such as \ce{H_{2}S} - are all known shock tracers \citep{SiOShockTracer,sulphurShockTracer} as a result of shock-driven evaporation of grain surface and/or mantle. Subsequent reaction of the evaporated species in the gas-phase, for example \ce{Si} and \ce{S} with \ce{O}, can produce molecules such as \ce{SiO} and \ce{SO}. Previous simulations \citep{SiOShockTracer,cjShocksFlower,UCLCHEM,jShocks} have used these molecules to great effect when tracing star formation related shocks under standard interstellar conditions.  However, as already demonstrated the CND and its surroundings represent extreme departures from standard interstellar conditions. For example, the OB star population in the Central Cavity could increase the photoionisation rate substantially. In addition to this, the cosmic-ray ionisation rate within the Central Molecular Zone is much larger than in other regions of the galaxy. Both \citet{sgrACRIRUL} and \citet{sgrACRIRLL} estimate this could extend to an upper limit of between $10 - 10^{4}\zeta$ \si{\per\second} assuming a standard galactic ionisation rate of $\zeta \approx 3 \times 10^{-17}$ \si{\per\second}. 
 
Additionally, the Central Molecular Zone and its immediate surroundings are known to exhibit large scale distributions of molecules such as \ce{CH_{3}OH} \citep{GCRMethanol} as well as \ce{HCN}, \ce{HNC} and \ce{HCO+} emission \citep{CMZLineSurvey}. Detection of these molecules, specifically \ce{CH_{3}OH}, in abundance is again typical of regions that have undergone grain processing. As noted by \citet{ch3ohFormation}, the gas-phase formation routes and efficiencies of \ce{CH_{3}OH} in particular cannot account for its measured abundance within the ISM. Instead, \citet{ch3ohFormation} found that \ce{CH_{3}OH} forms readily via successive Hydrogenation of \ce{CO} on the grain surface. Subsequent desorption - UV photodesorption being the dominant mechanism in translucent clouds - then releases \ce{CH_{3}OH} in to the gas-phase \citep{ch3ohFormation}. Crucially, these formation and desorption pathways are consistent with measured gas-phase abundances of \ce{CH_{3}OH}. \citet{galacticCenterCosmicRays} also investigated this phenomenon by comparing chemical models to \ce{CH_{3}OH} observations to constrain the mechanism responsible for its gas-phase existence. As \citet{galCenDustTemp} highlight, the dust temperature $\mathrm{T_{dust}} \leq 40$ \si{\kelvin} within the Central Molecular Zone is too low for thermal desorption to be significant. They therefore conclude that cosmic-ray desorption represents the dominant desorption mechanism of \ce{CH_{3}OH} within the Central Molecular Zone. However, on smaller scales applied to the CND both \citet{galCenKuiper} and \citet{Lau_2013} find that the dust temperature can vary positionally from 40 \si{\kelvin} -- 200 \si{\kelvin}, in turn showing that thermal desorption could become significant in sub-regions of the CND where the dust temperature exceeds $\approx$ 100 \si{\kelvin}. Consequently, detections of molecules such as \ce{SiO}, \ce{SO} and \ce{CH_{3}OH} are direct tracers of underlying physical events such \textbf{as} shocks, and/or regions that have undergone grain processing. 

Broadly, the detected emission intensity of these molecules is a function of the region's physical gas conditions which are, in turn, a function of the physical processes occurring within the gas. One may derive estimates of the kinetic gas temperature T$_\mathrm{kin}$, Hydrogen number density n$_{\mathrm{H}}$ and molecular column densities N$_{\mathrm{spec}}$ using radiative transfer codes such as RADEX \citep{radex} in order to understand these processes in more detail. Crucially, repeating this procedure for each position within and throughout a sample of beam pointings allows mapping of the gas conditions across those pointings, permitting a larger scale understanding of the physics and chemistry driving an object's evolution. 

Within this paper we present 0.38'' $\times$ 0.31'' spectral observations of Sgr A*'s Circumnuclear Disk, and apply the aforementioned radiative transfer methodology to this data. We present these observations in Section \ref{sec:obs} before discussing their implications in Section \ref{sec:obs-discussion}. In Section \ref{sec:sim} we outline our statistical process coupled to radiative transfer that infers the physical gas conditions towards each position. Sections \ref{sec:corner-plots} and \ref{sec:CND-gas-cond} presents the results of our modelling for each considered parameter, before Section \ref{sec:emission} discusses the possible emission mechanisms responsible for the observed complexity as well as our mass estimates.

\section{Observations} \label{sec:obs}

% \floattable
\begin{deluxetable}{ccc}
    \tablecaption{Details of the observational pointings used to collect data for this study. Shown are the Field ID for each pointing, alongside its central RA and Dec and the associated sensitivity. To assist future reference, Field IDs are prepended to numerical positions where molecular emission is identified, e.g. N1, N2 etc. Each pointing has size $\sim$ 21$^{\prime\prime}$ at 300 \si{\giga\hertz}. Spatial coverage for each pointing is shown in Figure \ref{fig:pointings}. \label{tab:pointings}}
    \tablewidth{0pt}
    \tablehead{
    Field & Pointing Center & Sensitivity \\
    ID & $\alpha$, $\delta$ (J2000)  & (mJy beam$^{-1}$ chan$^{-1}$)
    }
    \startdata
        N   & $17^h45^m40.60^s$,  $-29^\circ 00^{\prime} 20.00^{\prime\prime} $ & 4.98 \\
        G   & $17^h45^m43.80^s$,  $-29^\circ 00^{\prime} 20.00^{\prime\prime} $ & 5.24 \\
        D   & $17^h45^m41.75^s$,  $-28^\circ 59^{\prime} 47.69^{\prime\prime} $ & 5.34 \\
        H   & $17^h45^m39.49^s$,  $-28^\circ 59^{\prime} 51.02^{\prime\prime} $ & 5.02 \\
        J   & $17^h45^m38.00^s$,  $-29^\circ 00^{\prime} 01.00^{\prime\prime} $ & 5.24 \\
        K   & $17^h45^m38.30^s$,  $-29^\circ 00^{\prime} 40.00^{\prime\prime} $ & 5.35 \\
        L   & $17^h45^m39.95^s$,  $-29^\circ 01^{\prime} 03.50^{\prime\prime} $ & 5.46 \\
        M   & $17^h45^m41.32^s$,  $-29^\circ 01^{\prime} 01.63^{\prime\prime} $ & 5.39 \\
    \enddata
\end{deluxetable}

\begin{figure*}
    \centering
    \includegraphics[scale=0.48]{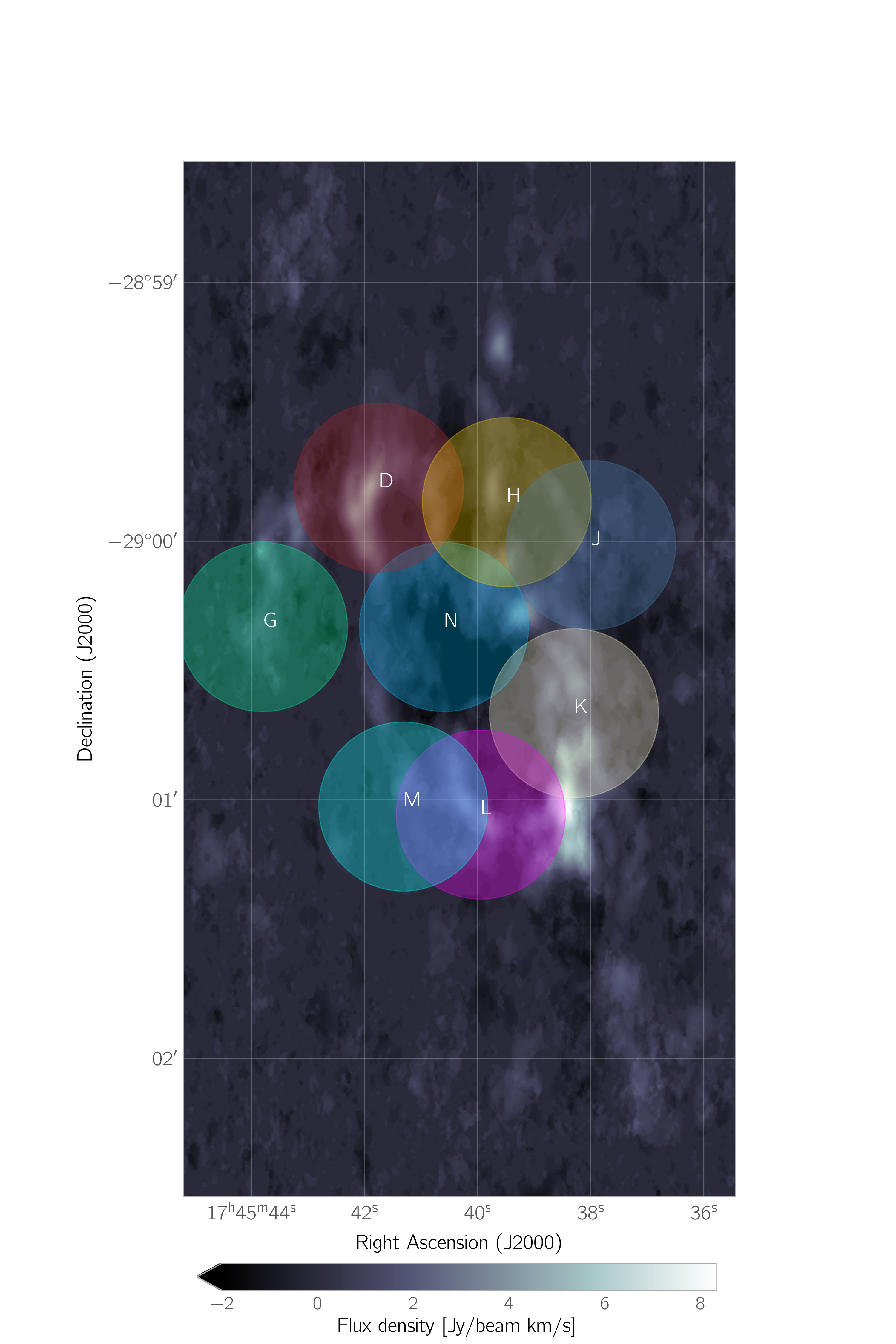}
    \caption{The 8 ALMA pointings of size 21$^{\prime\prime}$ at 300 \si{\giga\hertz} that comprise our data, overplotted on an \ce{HCN} emission map of Sgr A* and its surrounding Circumnuclear disk (CND). Details of the individual pointings can be found in Table \ref{tab:pointings}. We follow the naming convention in \citet{Christopher2005}, as well as Figure 2 of \citet{sgrANamingConvention}, for referring to regions within and around the CND, e.g. pointing G targets the Northeastern Arm whilst pointings M and L target the Southern Extension. Note: the G pointing is distinct and separate from the dust enshrouded sources of G1 and G2 in \citet{dustEnshroudedSources}.}
    \label{fig:pointings}
\end{figure*}

ALMA Cycle 2 Band 7 (272 - 375 \si{\giga\hertz}) observations (Project Code 2013.1.01242.S) that target various \ce{SiO} transitions towards the CND are utilized for this investigation.  The data presented in Figure \ref{fig:pointings} and Table \ref{tab:pointings} was obtained on 2015 June 8 and consists of 8 pointings with the 12m array. Each pointing has a total field of view of 19.608$^{\prime\prime}$ and a synthesized beam of 0.38$^{\prime\prime}$ $\times$ 0.31$^{\prime\prime}$ with PA$=-72^\circ$ corresponding to an on sky spatial scale of approximately $0.014 \times 0.012$ pc. We thus assume a beam filling factor of unity. The total integration time for each source is 151 \si{\second} with the relevant baseline tuned at the rest frequency of \ce{SiO (7 - 6)} (303.926809 \si{\giga\hertz}) along with a total bandwidth of 937.5 \si{\mega\hertz} corresponding to $\sim$~900 km s$^{-1}$.  The raw data product has a velocity resolution of $\sim$~0.5 \si{\kilo\meter\per\second} (3840 channels).

The data was calibrated with the assistance of standard observatory scripts using CASA 4.7.0 \citep{casa}.  CASA was also used to image the data with a 0.05$^{\prime\prime}$ pixel size and 2 \si{\kilo\meter\per\second} (2.03 Hz) channel width.  A Brigg's robustness parameter of 0.5 was utilized. The result achieved a sensitivity of $\sim$5 mJy, as shown in Table \ref{tab:pointings}.  Integrated and peak intensity maps were used to visually identify molecular line emission in the data cubes, and these identified lines are listed in Table \ref{tab:data-table} within Appendix \ref{sec:data-table}.

\subsection{CND spectra}

Spatial representations of the 8 pointings that comprise our observations are shown in Figure \ref{fig:pointings}. The spectra for each of the 99 positions is shown in Appendix \ref{sec:spectra}, whilst the spectra derived quantities are shown in Table \ref{tab:data-table} of Appendix \ref{sec:data-table}. The circular regions shown in Figure \ref{fig:pointings} represent the approximate field of view of each pointing corresponding to 21$^{\prime\prime}$ at 300 \si{\giga\hertz}.  Within this, we follow the naming convention adopted in \citet{Christopher2005} as well as \citet{sgrANamingConvention} for surveying regions of the CND. As such, the G pointing surveys the Northeastern Arm of Sgr A*, whilst all other pointings survey the interior of the CND. We note that the G-pointing and its sources within are distinct and separate from the dust enshrouded sources G1 and G2 highlighted by \citet{dustEnshroudedSources}. Within these pointings, each position contains at least either an \ce{SiO (7 - 6)} or \ce{SO (8_{7} - 7_{6})} line. Only 2 observations - N3, IRS7 - contain only \ce{SiO (7 - 6)} and \ce{SO (8_{7} - 7_{6})} respectively. 96 observations within the data contain both \ce{SiO (7 - 6)} and \ce{SO (8_{7} - 7_{6})} transitions; the exception is the M4 source in which we do not detect \ce{SiO (7 - 6)} but do detect both \ce{SO (8_{7} - 7_{6})} and \ce{CH_{3}OH (2_{1,1} - 1_{0,2})} transitions.

Whilst \ce{SiO} and \ce{SO} are the most commonly detected molecules, \ce{CH_{3}OH} forms the third most commonly detected species, being present towards 16 different positions. \ce{CH_{3}OH (2_{1,1} - 1_{0,2})} is not detected in any L pointing. In fact, it is most commonly observed within the N pointing. Additionally, our detections of \ce{H_{2}CS (9_{1,9} - 8_{1,8})} are only found within a small grouping of observations within the G pointing. Furthermore, we find \ce{OCS (25 - 24)} to have even further localisation, being present in only two sources: G2 and K13a. Curiously, G2 and K13a are located in two opposite, distinct regions within the CND and do not appear to be causally or spatially related to one another.

\section{Qualitative gas characteristics} \label{sec:obs-discussion}

\begin{figure*}
    \centering
    \includegraphics[scale=0.47]{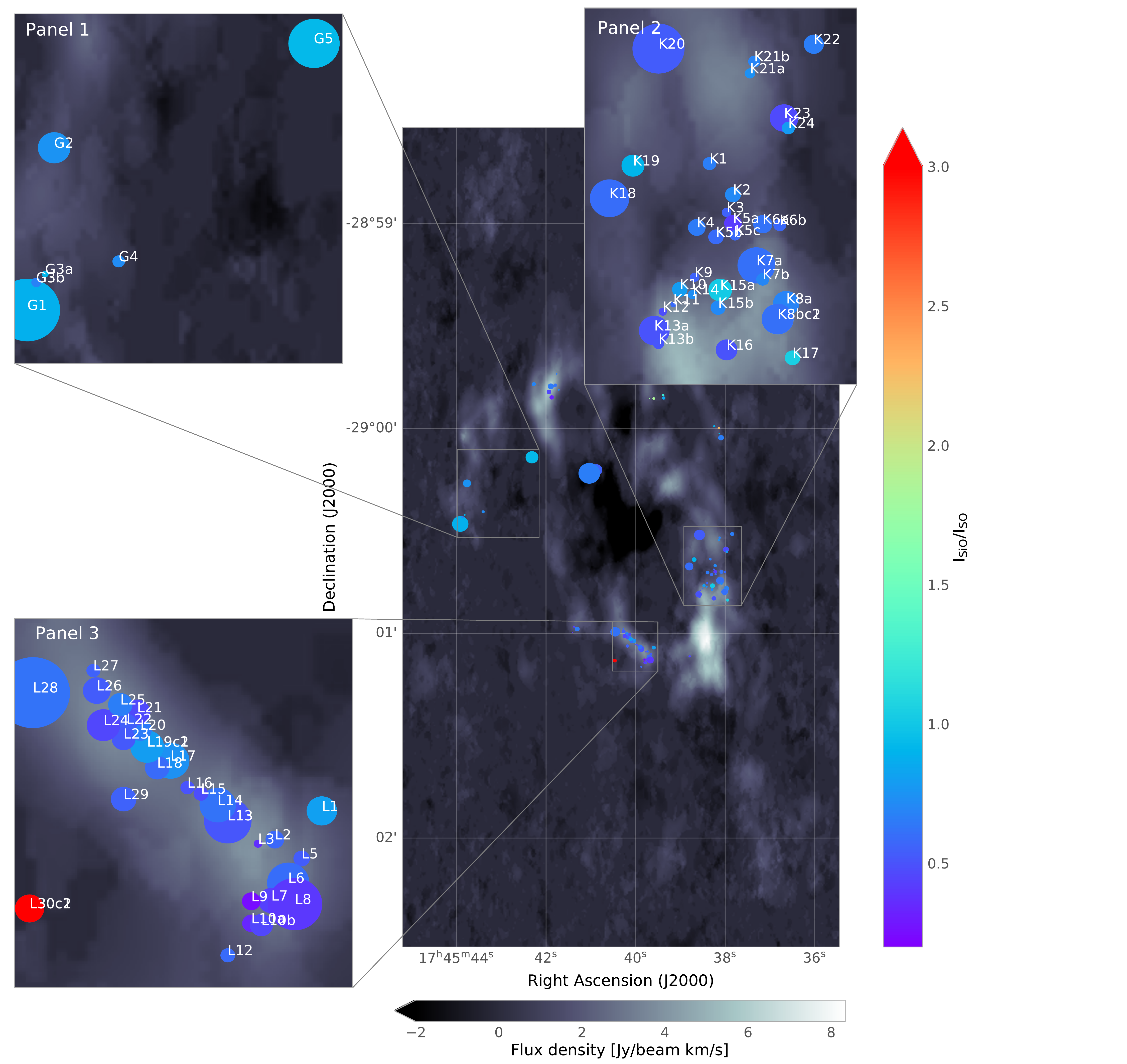}
    \caption{The ratio of observed \ce{SiO} to \ce{SO} flux densities $\mathrm{I_{SiO}/I_{SO}}$ for each position within the 8 pointings where \ce{SiO} and \ce{SO} are conclusively identified. Each circle represents the extent of each position in approximate synthesised beam sizes, and its colour denotes the ratio $\mathrm{I_{\ce{SiO}}/I_{\ce{SO}}}$ towards that region as per the vertical colour bar. Inset plots are labelled as Panel 1 -- 3 and show groupings of observations and their observational designations. Key here is the uniformity in $\mathrm{I_{\ce{SiO}}/I_{\ce{SO}}}$ throughout the dataset; most positions show $\mathrm{I_{\ce{SiO}}/I_{\ce{SO}}} < 1.5$.}
    \label{fig:sio-so-ratio}
\end{figure*}

Given the commonality of \ce{SiO} and \ce{SO} emission, the ratio $\mathrm{I}_{\ce{SiO}}/\mathrm{I}_{\ce{SO}}$ is a useful diagnostic of the most intense emission regions. Figure \ref{fig:sio-so-ratio} shows this ratio as a function of each position's location and extent. Each aperture within this plot represents the on-sky location of a detection described in Section \ref{sec:obs}, with the size of the aperture being approximately equivalent to the total extent in beams as given in Appendix \ref{sec:data-table}. The colour scale employed to fill these apertures is representative of the $\mathrm{I}_{\ce{SiO}}/\mathrm{I}_{\ce{SO}}$ ratio as indicated by the vertical colour bar. The underlying image represents \ce{HCN} emission of the CND. It's flux density is defined according to the horizontal colour bar.

Within Panel 1, the observations G1, G2, G3a and G3b sit along a filament of \ce{HCN} emission, whilst G4 and G5 are situated outside of this structure. As noted previously, we highlight that these positions and separate and distinct from the G1 and G2 sources identified in \citet{dustEnshroudedSources}. However, the presence of the filament appears to have negligible effect on the $\mathrm{I}_{\ce{SiO}}/\mathrm{I}_{\ce{SO}}$ ratio as all sources $< 1$. G1 is found to have a low $\mathrm{I}_{\ce{SiO}}/\mathrm{I}_{\ce{SO}}$ ratio of $\approx 0.8$, whilst G2 is observed to be slightly lower at $\approx 0.75$. The ratio at G3a is consistent with G1, whilst G3b is found to be slightly lower again at $\approx 0.7$. However, G4 has a ratio approximately consistent with G2, whilst G5 has ratio consistent with that of G1. Nevertheless, the $\mathrm{I}_{\ce{SiO}}/\mathrm{I}_{\ce{SO}}$ ratio implies that the gas in this region has marginally greater \ce{SO} emission than \ce{SiO}, in turn indicating a higher density of \ce{SO} than \ce{SiO} within this region. 

When considering Panel 2, it is immediately evident that there is again very little variation in the $\mathrm{I}_{\ce{SiO}}/\mathrm{I}_{\ce{SO}}$ ratio across the area covered. With the exception of K15a and K17, every other observation in this inset plot shows $\mathrm{I}_{\ce{SiO}}/\mathrm{I}_{\ce{SO}} \lessapprox 1$. Interestingly, K12, K13a and K13b sit along a ridge that demarcates a bright \ce{HCN} emission cloud from its surrounding background. It is this region that is subject to close interaction between the Galactic Center Mini Spiral and the molecular gas within the CND. From the $\mathrm{I}_{\ce{SiO}}/\mathrm{I}_{\ce{SO}}$ ratio, it would appear that this ridge shows stronger \ce{SO} emission than \ce{SiO} emission. 

Panel 3, much like Panel 2, shows very little variation in ratio. The only exception to this uniformity is the multi-component gas at L19 and L30. L19's surrounding gas shows only single component behaviour, making the multi-component nature of L19 puzzling. Unfortunately, L30 does not have any nearby observation to compare with, though it is worth noting than the $\mathrm{I}_{\ce{SiO}}/\mathrm{I}_{\ce{SO}}$ ratio for the second component here is the largest ratio seen in the data at $\approx 12$. This stands in stark contrast to the rest of the observations in Panel 3 whose ratios are significantly lower and much more uniform at $\lessapprox 1$. 

Moreover, in addition to L19 and L30 we find 2 further sources in our observations - D3, and K8b - that exhibit multi-component behaviour. Curiously, each of these sources contains only \ce{SiO} and \ce{SO} lines. 

Broadly, Figure \ref{fig:sio-so-ratio} seemingly indicates that the energetics of the CND do not vary significantly throughout its extent, or from region to region. However, $\mathrm{I}_{\ce{SiO}}/\mathrm{I}_{\ce{SO}}$ is inherently degenerate as the ratio when  $\mathrm{I}_{\ce{SO}} << 1$ can also be the same as when $\mathrm{I}_{\ce{SiO}} >> 1$. We therefore utilise a quantitative approach to deriving the gas conditions within each observation.

\section{Gas condition inference procedure} \label{sec:sim}

Both the physics and chemistry of interstellar gas are entangled and thus form a system that is extremely non-linear and non-trivial to model. We simplify this modelling challenge by approaching the problem with non-LTE radiative transfer in RADEX \citep{radex} coupled to a Bayesian inference procedure. Within this scheme we may inform the radiative transfer with a set of independent random variables that define the physical gas conditions. Monte Carlo methods can then be used to explore the physical condition parameter space, in turn converging on the most likely gas conditions for each region. Additionally, this approach minimises the number of variables required to model the CND gas, in turn allowing us to determine the physical gas conditions with a minimal number of assumptions. A key assumption for applying RADEX is, however, that the transitions under consideration are optically thin, as is likely the case.

\subsection{Radiative transfer} \label{sec:radiative-transfer}

Using this approach, we utilise the radiative transfer code RADEX \citep{radex} to compute the emitted line flux density of the relevant transition for a parcel of gas with physical conditions described by the set of parameters $\mathcal{S} = \{\mathrm{T}_{\mathrm{kin}}, \: n_{\mathrm{H}}, \: N_{\mathrm{spec}} \}$. T$_{\mathrm{kin}}$ is the gas temperature, n$_{\mathrm{H}}$ is the molecular Hydrogen gas number density and N$_{\mathrm{spec}}$ is the column density of the relevant species under consideration. Owing to the strong UV heating supplied by the OB star population, as well as the high temperatures inf erred in the CND from previous observations, we explore temperatures up to 1000 \si{\kelvin}. We used RADEX owing to its non-LTE molecular excitation capabilities, as well as its widespread use in studies that model interstellar spectra \citep[e.g.][]{chemDiversity,nonLTENO,wishWaterModelling}, including towards the Galactic Center \citep[e.g.][]{sgrB2Ammonia,sgrAPDR,sgrB2ShockedGas}. We intend to apply more detailed chemical modelling to these regions in future in order to further understand their phenomenology. All species within this study utilise temperature dependent collisional data from the LAMDA database \citep{lamda}. For temperatures outside of the calculated temperature range that LAMDA reports, RADEX fixes the collisional rate constant to the rate constant at the nearest temperature. For \ce{SiO}, the collisional data has been calculated up to and including collisional temperatures of 1000 \si{\kelvin}. Inspecting the rate constants for \ce{SiO (8_{7} - 7_{6})} reveals relatively uniform (within a factor of 2) behaviour across the a-priori temperature range. Considering a similar process for \ce{SO (7 - 6)} we note that the collisional rates have been determined up to 300 \si{\kelvin}. Much like \ce{SiO}, the collisional data for \ce{SO} is similarly constant within the calculated temperature range. We therefore assume that, given the similar chemical compositions and masses of \ce{SiO} and \ce{SO}, the same uniformity in collisional rates is evident beyond 300 \si{\kelvin}. We note similar behaviour within the collisional data for the other molecular transitions detected in our data, and we therefore apply the same assumption to all molecules and transitions detected in our observations.

Additionally, each observation presented in Section \ref{sec:obs} has been normalised to units of mJy$/$beam and thus assumes that each observation fills the beam. This renders RADEX computed flux densities, as RADEX assumes a beam filling factor of $1$, directly comparable to the observational flux densities presented. The assumption of a beam filling factor of unity is further justified as the linear scale of our observations is sufficient to guarantee that the only emission within the beam is from the targeted gas. Each source is also fully resolved, in turn ensuring they fill each beam.

Repeating this procedure for each species and each transition in a given observation allows us to construct a model $\boldsymbol{\theta}$ that represents the synthetic flux density emitted from a region of gas at temperature T$_{\mathrm{kin}}$ and gas density n$_{\mathrm{H}}$ comprising species column densities N$_{\mathrm{spec}}$. Constructing the model in this manner assumes that all emission lines contained within it originate from the same thermal component. As Section \ref{sec:obs} showed, our observations are consistent with a single component. Where not, it is clear and distinct that $2$ components are present, either through distinctly different linewidths or bi-modal line profiles, and they are thus treated as essentially two separate sources, in turn being analysed with $2$ distinct models. 

However, given the complexity of the underlying physics and chemistry, and indeed their associated degeneracy, it is useful to statistically survey the entire parameter space across each dimension in order to better understand and identify the probable conditions within each observation. An ideal process with which to do this utilises Bayesian Inference. 

\subsection{Bayesian Inference}

Bayesian Inference is a method of using conditional probability to build statistical parameter distributions for multi-dimensional data. Unsurprisingly, it utilises Bayes' Theorem, defined in Equation \ref{eq:bayes}.

\begin{equation}
    P(\boldsymbol{\theta} | \boldsymbol{d}) = \frac{P(\boldsymbol{d} | \boldsymbol{\theta}) P(\boldsymbol{\theta})}{P(\boldsymbol{d})} \label{eq:bayes}
\end{equation} 

$P(\boldsymbol{\theta} | \boldsymbol{d})$ is defined as the probability of the model $\boldsymbol{\theta}$ fitting the data $\boldsymbol{d}$ given that the data is correct (known as ``the posterior probability''). $P(\boldsymbol{d} | \boldsymbol{\theta})$ is the probability of having obtained the data $\boldsymbol{d}$ given a particular model $\boldsymbol{\theta}$ (known as ``the likelihood''). $P(\boldsymbol{\theta})$ is defined as the probability of the model $\boldsymbol{\theta}$ being correct (``the prior'') and $P(\boldsymbol{d})$ is the probability that the data is correct (``the evidence''). We define the data as being the observed flux densities, whilst the model is the equivalent line flux density determined from RADEX as defined in Section \ref{sec:radiative-transfer}.

We assume flat priors in log space such that within each dimension, any value is equally as likely as any other providing that such a value falls within the non-zero prior range. The non-zero ranges used for these priors, and in turn the extent of parameter spaces explored, are shown in Table \ref{tab:params}. 

\begin{deluxetable}{l l l}[htbp!]
    \tablecaption{The pre-constrained parameter spaces surveyed in this preliminary study for each explored dimension. T$_{\mathrm{kin}}$ represents the gas kinetic temperature, n$_{\mathrm{H}}$ is the Hydrogen gas number density and N$_{\mathrm{spec}}$ is the species column density corresponding to the species $\mathrm{spec}$. \label{tab:params}}
    \tablehead{\colhead{Parameter}  & \colhead{Range}  & \colhead{Unit} }
    \startdata
    T$_{\mathrm{kin}}$  & $60$-$1000$ & \si{\kelvin} \\ 
    n$_{\mathrm{H}}$ & $10^{3}$-$10^{7}$ & \si{\per\centi\meter\cubed}  \\ 
    N$_{\mathrm{SiO}}$ &  $10^{12}$-$10^{16}$ & \si{\per\centi\meter\squared}      \\ 
    N$_{\mathrm{SO}}$ &  $10^{12}$-$10^{16}$ & \si{\per\centi\meter\squared}     \\ 
    N$_{\mathrm{CH_{3}OH}}$ & $10^{14}$-$10^{17}$ & \si{\per\centi\meter\squared}     \\ 
    N$_{\mathrm{H_{2}CS}}$ & $10^{12}$-$10^{15}$ & \si{\per\centi\meter\squared}     \\
    N$_{\mathrm{OCS}}$ & $10^{12}$-$10^{15}$ & \si{\per\centi\meter\squared}     \\ 
    \enddata
\end{deluxetable}

We constrain the upper and lower limits of column density for each molecule to explore by estimating the LTE derived column density for the upper and lower limits of temperature shown in Table \ref{tab:params}. This makes use of molecular data provided by Splatalogue \citep{splatalogue}.

We compute the likelihood using the likelihood function defined in Equation \ref{eq:likelihood}.

\begin{equation} 
    P(\boldsymbol{d} | \boldsymbol{\theta}) = \exp\Bigg(-\frac{1}{2} \sum_{i} \frac{(d_{i} - \theta_{i})^{2}}{\sigma_{i}^{2}} \Bigg) \label{eq:likelihood}
\end{equation} 

Where the summation term represents the Chi-squared statistic. Within Equation \ref{eq:likelihood}, $d_{i}$ is the i\textsuperscript{th} value of the data, $\theta_{i}$ is the i\textsuperscript{th} value of the model corresponding to $d_{i}$ and $\sigma_{i}$ is the error associated with $d_{i}$. 

By implementing this Bayesian procedure we sample from the underlying posterior distribution using randomly sampled variables that define the gas condition. These variables thus form the set $\mathcal{S}$. To perform this sampling we use the affine-invariant MCMC sampler \texttt{emcee} \citep{emcee}. \texttt{emcee} uses a numbers of ``walkers'' to systematically sample the parameter space across a number of steps, in turn constructing a series of Markov Chains. All walkers therefore define a representation of the underlying posterior distribution, allowing us to identify statistically informed estimates of the most likely gas condition towards each observation. We use an ensemble of $500$ walkers and we run each walker for a maximum of $1\times10^{4}$ steps. Every $100$ steps we assess convergence in each dimension using the Gelman-Rubin statistic \citep{gelmanRubin}. We halt our sampling if convergence is judged to have been achieved across all dimensions. We follow \citet{brooksAndGelman} in judging successful convergence when the Gelman-Rubin diagnostic for each dimension $< 1.2$. For additional confidence, we select a cut-off of $1.15$ for achieving sample convergence. In addition and where relevant we also visualise our chains using \texttt{ChainConsumer} \citep{chainConsumer}.

This approach has a number of distinct advantages when deriving physical gas conditions. Firstly, the number of variables required to define the gas is limited to those required by RADEX, resulting in a relatively low number of dimensions overall. Additionally, within this scheme each dimension is entirely free and is not causally related to any other. As such, we derive the gas conditions with almost no a-priori knowledge of the local chemistry. The affine-invariant nature of the \texttt{emcee} sampler thus allows convergence to be achieved independently of the underlying physics and/or chemistry beyond radiative transfer, thus reducing uncertainty and bias.

\section{Inference results: corner plots for individual sources} \label{sec:corner-plots}

\begin{figure*}
    \centering
    \includegraphics[scale=0.265]{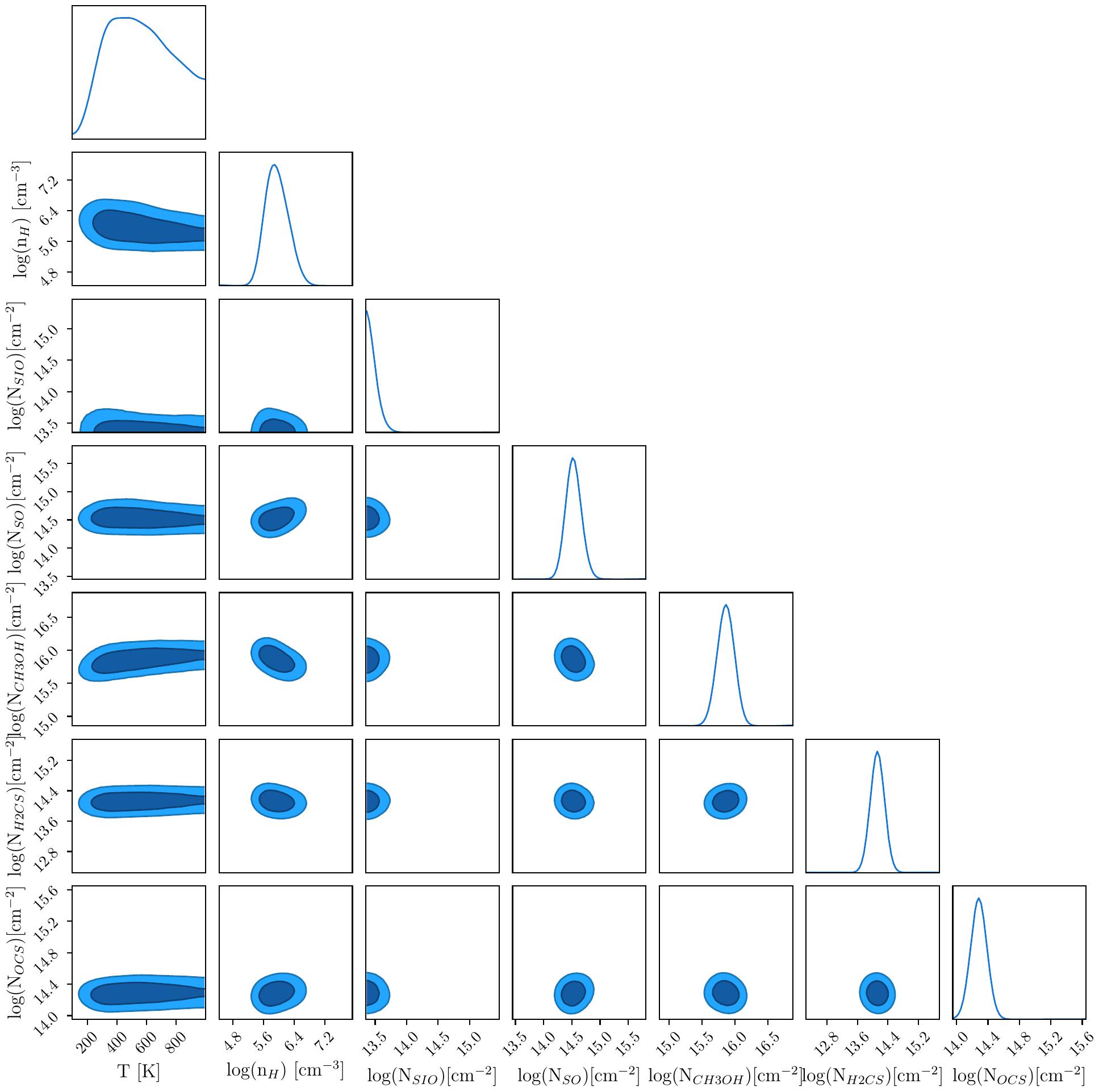}
    \caption{An example corner plot for the G2 source that shows the sampled distributions for each parameter, known as dimensions, in the model. Each dimension is shown on the $x$ axis; the distributions on the diagonal represents the posterior distributions for the relevant dimension e.g. the posterior on the top left is the $\mathrm{T}$ posterior. All other plots in the matrix show the joint distributions of parameters as per their $x$ and $y$ axes that contribute to the associated posterior. Tight constraints are observed in the joint distributions within all dimensions except $\mathrm{T}$ which exhibits broad behaviour along its domain and highlights the broad errors associated with this dimension.}
    \label{fig:mcmc-G2}
\end{figure*}

Figure \ref{fig:mcmc-G2} shows an example of the corner plot for the G2 source. G2 contains the largest number of lines in our sample that our detected towards one region. 

Within this matrix, the marginalised posterior distributions for each parameter are found on the diagonal. The peak of each of these distributions represents the point of highest posterior density (HPD) and it this value that is taken to be the most likely model given the data. The plots found within the remaining cells represent the joint distributions for the parameters defined by their relevant $x$ and $y$ axes. Any pair of parameters that are bounded by a dark blue region is within $2\sigma$ of the HPD, whilst everything bounded within the light blue region is within $3\sigma$ of the HPD.

Considering the marginalised posterior distributions we see that each dimension peaks to a maximum, though not without a degree of skewness in the case of $\mathrm{T}$. Inspecting the joint distributions of all dimensions except $\mathrm{T}$ shows roughly symmetrical, tight groupings that in general extend to $\pm 0.5$ orders of magnitude. This produces posteriors that peak strongly, though both \ce{SiO} and \ce{OCS} peak towards the lower boundary of the distributions in this case. 

Whilst estimates of $\mathrm{T}$ are broad, its HPD is reliably constrained in the case of G2. However, the skewness in this distribution does serve to highlight the assymetrical errors that we find are often associated with this dimension. In this instance, the possible values $\mathrm{T} < 350$ \si{\kelvin} become exponentially less likely. At $\mathrm{T} > 350$ \si{\kelvin} we observe a small plateau of equal likelihood until $\mathrm{T} \approx 600$ \si{\kelvin}, whereby an approximately linear decrease in likelihood until $\mathrm{T} \approx 1000$ \si{\kelvin} is encountered. 

\begin{figure*}
    \centering
    \includegraphics[scale=0.3]{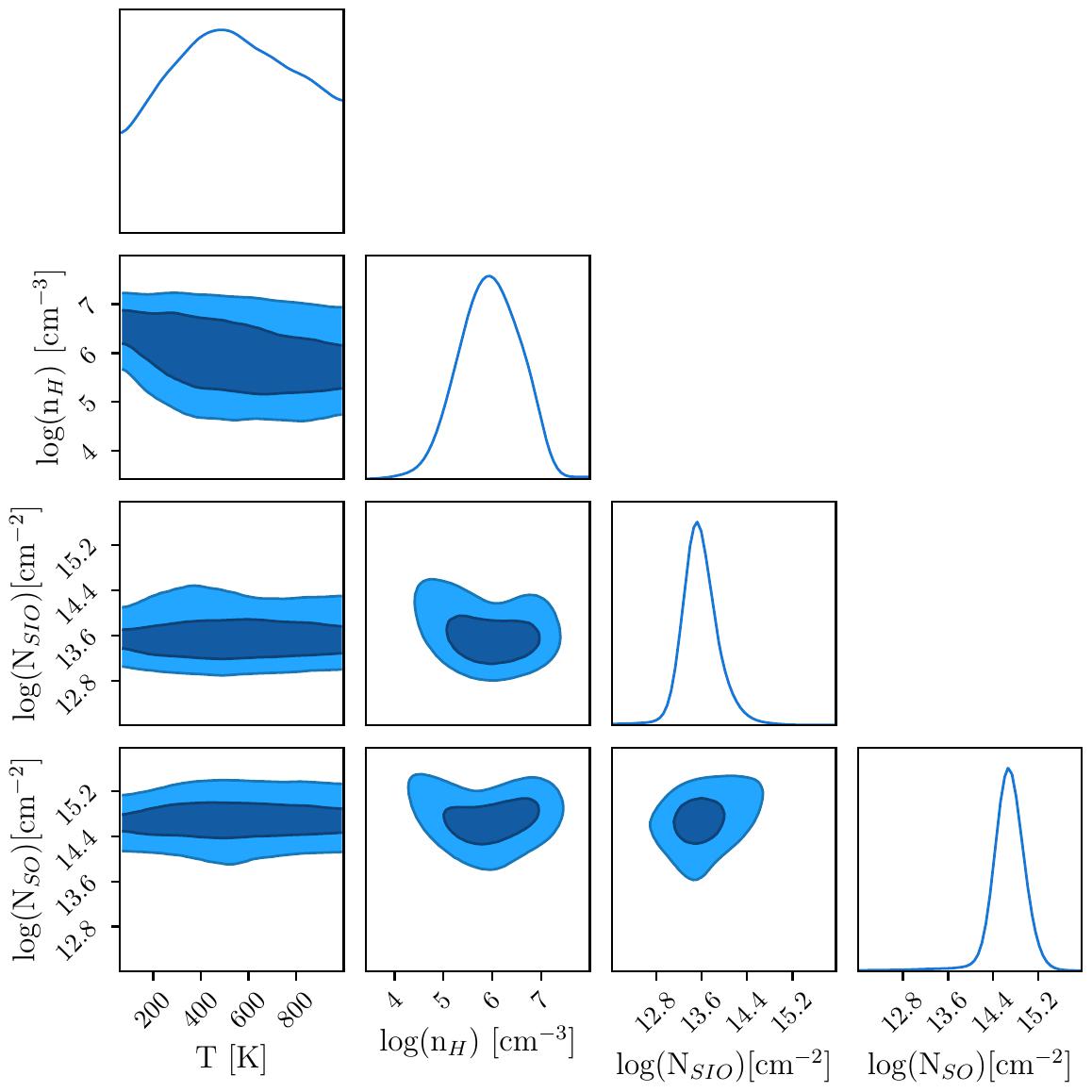}
    \caption{Corner plot for the K1 source as per Figure \ref{fig:mcmc-G2}. This shows much broader spread in all dimensions owing to the limited number of data points available. }
    \label{fig:mcmc-k1}
\end{figure*}

Unfortunately, the large errors and associated systematic uncertainties in temperature are a consequence of the limited number of spectral lines detected and used in this analysis. In the majority of cases within this study we require fitting for at least 4 different dimensions using just 2 spectral lines. An example corner plot for this scheme is shown in Figure \ref{fig:mcmc-k1}, which exhibits only \ce{SiO} and \ce{SO} lines. As a result, Figure \ref{fig:mcmc-k1} shows broader constraints in all dimensions and less symmetry in the posterior distributions therein, though $\mathrm{T}$ does peak despite $3\sigma$ errors spanning the entire parameter space. Subsequent inspection of the temperature column of Table \ref{tab:mle-conditions} shows wide, asymmetrical errors on almost every constrained temperature, serving to highlight the challenge in drawing a discrete conclusion from a continuous, asymmetrical and weakly peaked distribution. The errors on each parameter become both smaller and more symmetrical in nature, when afforded more line detections, though the effect on constraining $\mathrm{T}$ is slight. Additionally, the broad temperature distributions may be reflective of the weak temperature dependence that is observed in the LAMDA collisional data for the detected molecular transitions. Further constraints could be applied if additional observations of molecular transitions that exhibit strong temperature dependent behaviour are incorporated.

\section{Inference results: Gas conditions in the CND} \label{sec:CND-gas-cond}

The derived HPD of the physical parameters for each position are shown in Figures \ref{fig:CND-T} - \ref{fig:CND-N-so}. These plots are similar to that in Figure \ref{fig:sio-so-ratio}, however the colour scale employed to fill the apertures here is indicative of the HPD of the relevant dimension under consideration as computed from our inference routine. For instance, Figure \ref{fig:CND-T} shows the HPD of gas temperature T$_{\mathrm{kin}}$ towards each observation. Similarly, Figures \ref{fig:CND-n}, \ref{fig:CND-N-sio} and \ref{fig:CND-N-so} show the HPDs of n$_{\mathrm{H}}$, N$_{\mathrm{SiO}}$ and N$_{\mathrm{SO}}$ respectively. Semi-transparent, hatched apertures represent sources that have not achieved reliable convergence within the maximum number of steps available. We define non-reliable convergence as any source that has converged to an HPD but without an associated $1\sigma$ error range. Table \ref{tab:mle-conditions} within Appendix \ref{sec:mle-conditions} shows the HPD for each source and each dimension therein together with their associated error ranges. Thus considering each image allows the dataset wide distribution of physical conditions to be investigated and quantified.

\subsection{Temperature} \label{sec:CND-gas-cond-T}

\begin{figure*}
    \centering
    \includegraphics[scale=0.47]{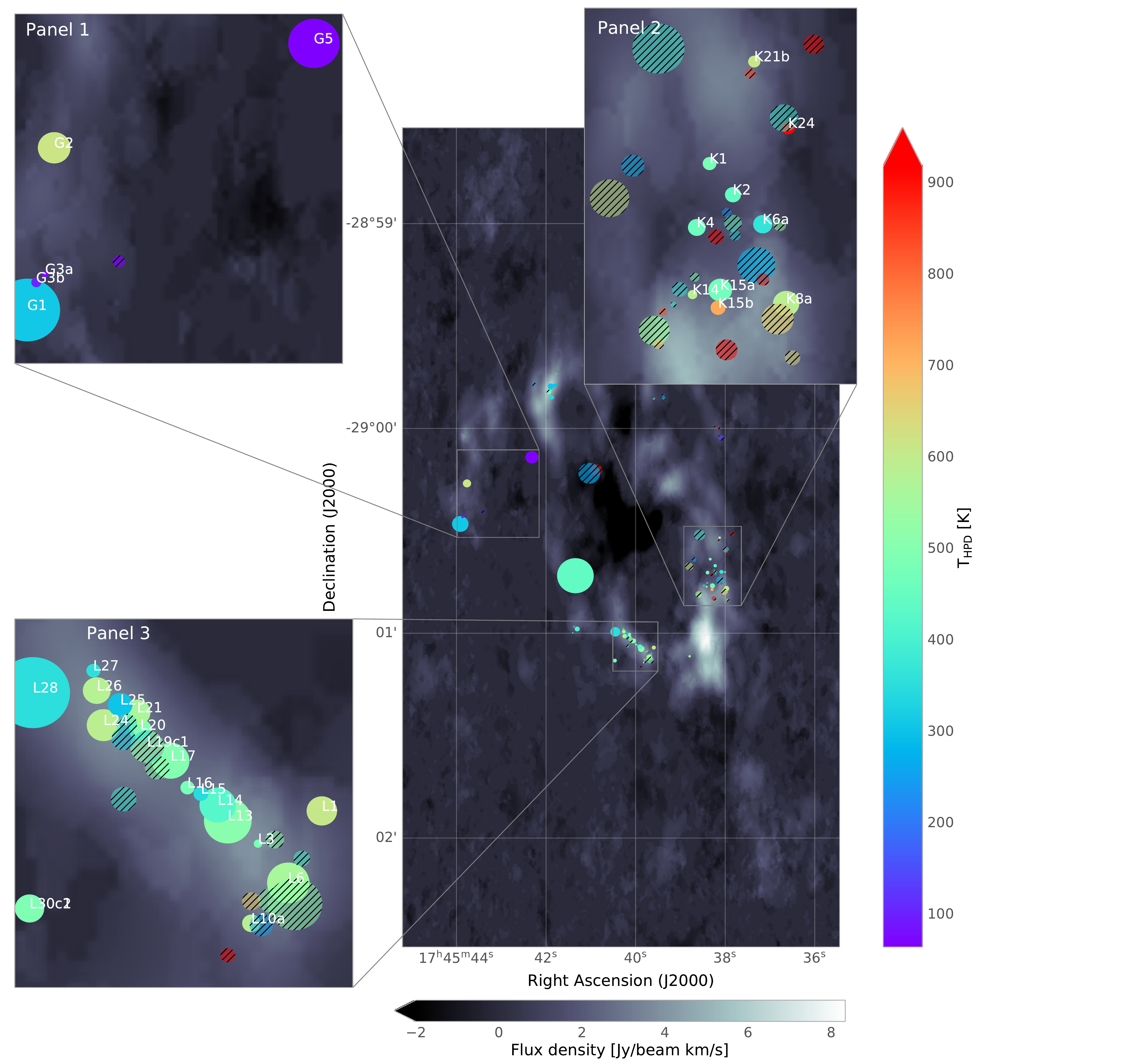}
    \caption{The points of highest posterior density (HPD) of the kinetic gas temperature T$_{\mathrm{HPD}}$ towards each position outlined in Section \ref{sec:obs}. Semi-transparent, hatched circular regions indicate a position whose chain has not reliably converged. G2 (in Panel 1) is noticeably warmer than its surroundings, which are much lower in their recovered temperatures. Panels 2 and 3 show more agreement between neighbouring positions.}
    \label{fig:CND-T}
\end{figure*}

The distribution of HPD temperatures T$_{\mathrm{HPD}}$ across the entirety of Figure \ref{fig:CND-T} shows minor correlation on a local level; as expected, sources in close proximity to one another have broadly similar temperatures. In general, this behaviour manifests as $\pm 50$ \si{\kelvin} between neighbouring sources. An example of this is found in Panel 1 which shows that G3a and G3b are almost identical in temperature of $\approx 100$ \si{\kelvin}. Both G3a and G3b, as well as G1 and G2, are found within a long filamentary structure of \ce{HCN} emission that extends along the left vertical edge of Panel 1. This structure can be regarded as sitting outside of the main CND structure, and represents the 50 \si{\kilo\meter\per\second} molecular cloud identified in previous studies. However, both G1 and G2 are warmer than G3a and G3b with T$_{\mathrm{G1}} \approx 350$ \si{\kelvin} and T$_{\mathrm{G2}} \approx 600$ \si{\kelvin}. Additionally, Panel 1 also shows G5, which is external to the \ce{HCN} filament. G5 is similar in temperature to G3a and G3b with T$_{\mathrm{G5}} \approx 100$ \si{\kelvin}. There is also an unconstrained source on the boundary of the filament that is of similar temperature to G3a, G3b and G5, though this source's temperature estimate is likely unreliable. Nevertheless, it would appear that the G3a and G3b observations capture the same gas component, whilst G1 and G2 capture colder and warmer extremes of the filament respectively. 

When considering Panel 2, we see that the main grouping of observations in the centre of the panel are consistent with temperatures $400 < \mathrm{T_{{HPD}}} < 700$ \si{\kelvin}, though there is a significant number of positions that have not achieved reliable temperature convergence. Additionally, $5$ of the converged positions are identified as having T$_{\mathrm{HPD}} \approx 500$ \si{\kelvin}. Interestingly, this is consistent with the bulk of the gas being close to the upper limit of the CND clump temperature as identified by \citet{cndClumpConditions}. There are, however, 2 constrained ``hot'' sources within Panel 2 that exceed this limit: K15b and K24. Of these, K24 is the hottest at T$_{\mathrm{K24}} \approx 800$ \si{\kelvin}, whilst K15b is moderately cooler at T$_{\mathrm{K15b}} \approx 700$ \si{\kelvin}. K24 is isolated and much further North than K15b, and does not appear to correlate with the underlying \ce{HCN} emission. Conversely, K15b sits embedded within the CND and is much closer to cooler sources. This therefore indicates that in general the gas temperature is increasing towards the South of Panel 2 and, in turn, to the South of the CND. This is also consistent with the underlying \ce{HCN} emission becoming brighter towards the South of Panel 2.

Panel 3 shows a similar trend to that observed in the main grouping of Panel 2 in that almost every source in Panel 3 has temperature within the range $350 < \mathrm{T_{{HPD}}} < 550 $ \si{\kelvin}. In this instance there are only 2 sources - L15 and L25 - that deviate significantly from $\approx 550$ \si{\kelvin}. L25 sits at the furthest extremity of the observation grouping, whilst L15 sits centrally within that grouping. However, we observe that T$_{\mathrm{HPD}}$ increases from L15 along and down the line of observations towards L6, with L3 and L6 being of comparable temperature. The $2$ sources either side of L6 - L10a and L1 - are again of comparable temperatures to L3 and L6. Continuing along and up from L15 towards L20 shows that the tight grouping of observations in L20, L21 and L24 are all of similar temperature, again T$_{\mathrm{HPD}} \approx 550$ \si{\kelvin}. However, L25's temperature of T$_{\mathrm{HPD}} \approx 350$ \si{\kelvin} stands in contrast to its warmer surroundings. 

Additionally, L19 is the observation that showed the largest $\mathrm{I}_{\ce{SiO}}/\mathrm{I}_{\ce{SO}}$ ratio in Panel 3 as well as globally across all observations. Panel 3 shows that its temperature is unconstrained, though if we consider the estimate in relation to its surroundings then we see no significant deviation from the average T$_{\mathrm{HPD}}$ of its surrounding observations.

\subsection{Gas density} \label{sec:CND-gas-cond-n}

\begin{figure*}
    \centering
    \includegraphics[scale=0.47]{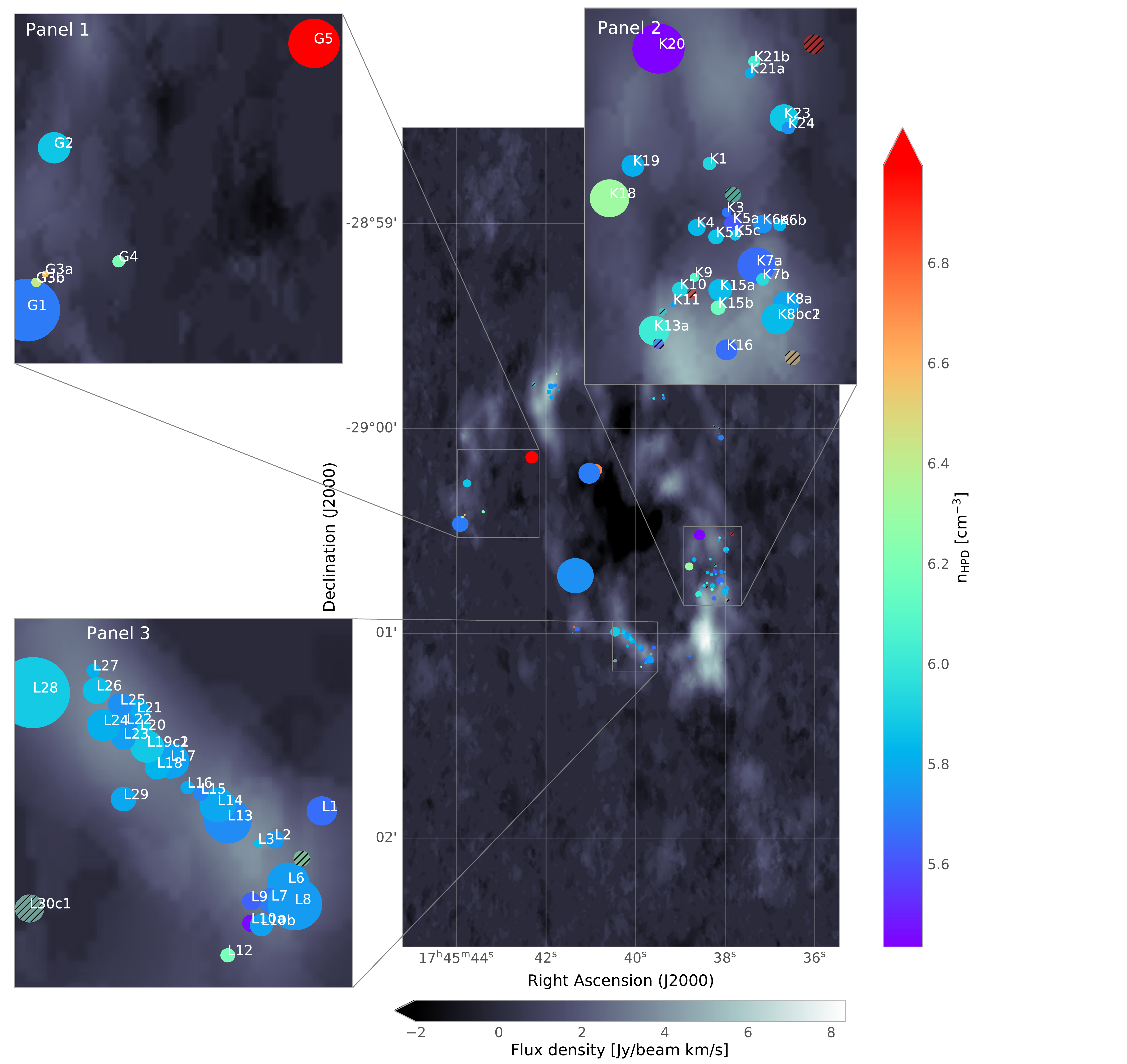}
    \caption{As per Figure \ref{fig:CND-T}, but here the colour of the circular regions represents Hydrogen gas density n$_{\mathrm{H}}$ towards each position outlined in Section \ref{sec:obs}. Again, semi-transparent, hatched circular regions indicate a source whose chain has not reliably converged.}
    \label{fig:CND-n}
\end{figure*}

When considering the distribution of HPD densities n$_{\mathrm{HPD}}$ in Figure \ref{fig:CND-n}, we see a similar local trend to that observed in temperature in Figure \ref{fig:CND-T} such that neighbouring sources are within $0.2$ orders of magnitude of one another. Additionally, we also observe far fewer unconstrained sources than were noted in Figure \ref{fig:CND-T}. However, more uniformity is seen in the determined values of n$_{\mathrm{HPD}}$ throughout the CND as a whole, with less than $2$ orders of magnitude variation throughout. 

Panel 1 shows that, in general, colder sources are found to be at higher density. For example, G5 was identified as being $\approx 100$ \si{\kelvin} and is, in turn, here noted to have density n$_{\mathrm{G5}} \approx 10^{7}$ \si{\per\centi\meter\cubed} - the most dense of any position in our sample. Whilst G1 and G2 were noted for their different temperatures (G1 becoming cooler than G2), Panel 1 shows that they are broadly similar in density within their associated error ranges. Furthermore the same density behaviour is seen in G3a, G3b and G4 which were all identified as having similar temperatures. 

When considering Panel 2, almost every source sits within the range $10^{5.6} < \mathrm{n_{HPD}} < 10^{6.2}$ \si{\per\centi\meter\cubed}, even those outside of the main grouping identified in Panel 2 of Figure \ref{fig:CND-T}. The main grouping is found within an \ce{HCN} bright region that defines the Southern extent of the CND. We note that the density range derived here is consistent with the density of \ce{HCN} bright cores estimated by \citet{sgrAHCN}. Additionally, the hot sources identified in K15b and K24 sit within this range and have densities n$_{\mathrm{K15b}} \approx 10^{6.2}$ \si{\per\centi\meter\cubed} and n$_{\mathrm{K24}} \approx 10^{5.8}$ \si{\per\centi\meter\cubed} respectively. K13a is identified as having a density towards the upper limit, i.e. $\mathrm{n_{HPD}} \approx 10^{6.2}$ \si{\per\centi\meter\cubed}, though its surroundings are of lesser density. When considered in conjunction with K9 and K10, it would appear that this region is tracing the dense inner-most ridge of the CND.

Outside of this grouping we identify that K18 exhibits a large density of $\mathrm{n_{HPD}} \approx 10^{6.4}$ \si{\per\centi\meter\cubed}. This is not consistent with the underlying \ce{HCN} emission and is not reflected in the density of its nearest neighbour K19 whose density is $\mathrm{n_{HPD}} \approx 10^{5.8}$ \si{\per\centi\meter\cubed}. K18 is therefore regarding as an anomalous position.

Within Panel 3 we observe that the L16 source acts as a dividing boundary such that everything North of L16 - and thus the Northern-most cluster of observations - all exist at almost exactly the same density of n$_{\mathrm{HPD}} \approx 10^{6.0}$ \si{\per\centi\meter\cubed}. South of L16 we note that the sources show slightly more variety, with L3 being more dense than the average of n$_{\mathrm{HPD}} > 10^{5.8}$ \si{\per\centi\meter\cubed} and L16 being less dense at n$_{\mathrm{HPD}} > 10^{5.5}$ \si{\per\centi\meter\cubed}. L12 is a unique outlier in that its density is significantly larger than any other source in Panel 3 at n$_{\mathrm{HPD}} > 10^{6.4}$ \si{\per\centi\meter\cubed}. Within Figure \ref{fig:CND-T}, L12 does not have a reliable temperature estimate. However, if we consider the temperature identified we observe that it is $> 800$ \si{\kelvin}, thus potentially forming another non-quiescent region. 

Additionally, observations that sit within \ce{HCN} regions or filaments are found to broadly exhibit higher densities than those observations outside of these locations. Given that \ce{HCN} emission traces dense gas this is unsurprising, though it does lend additional weight to the hot, dense sources being embedded non-quiescent regions such as a YSO.

\subsection{Column density} \label{sec:CND-gas-cond-N}

\begin{figure*}
    \centering
    \includegraphics[scale=0.47]{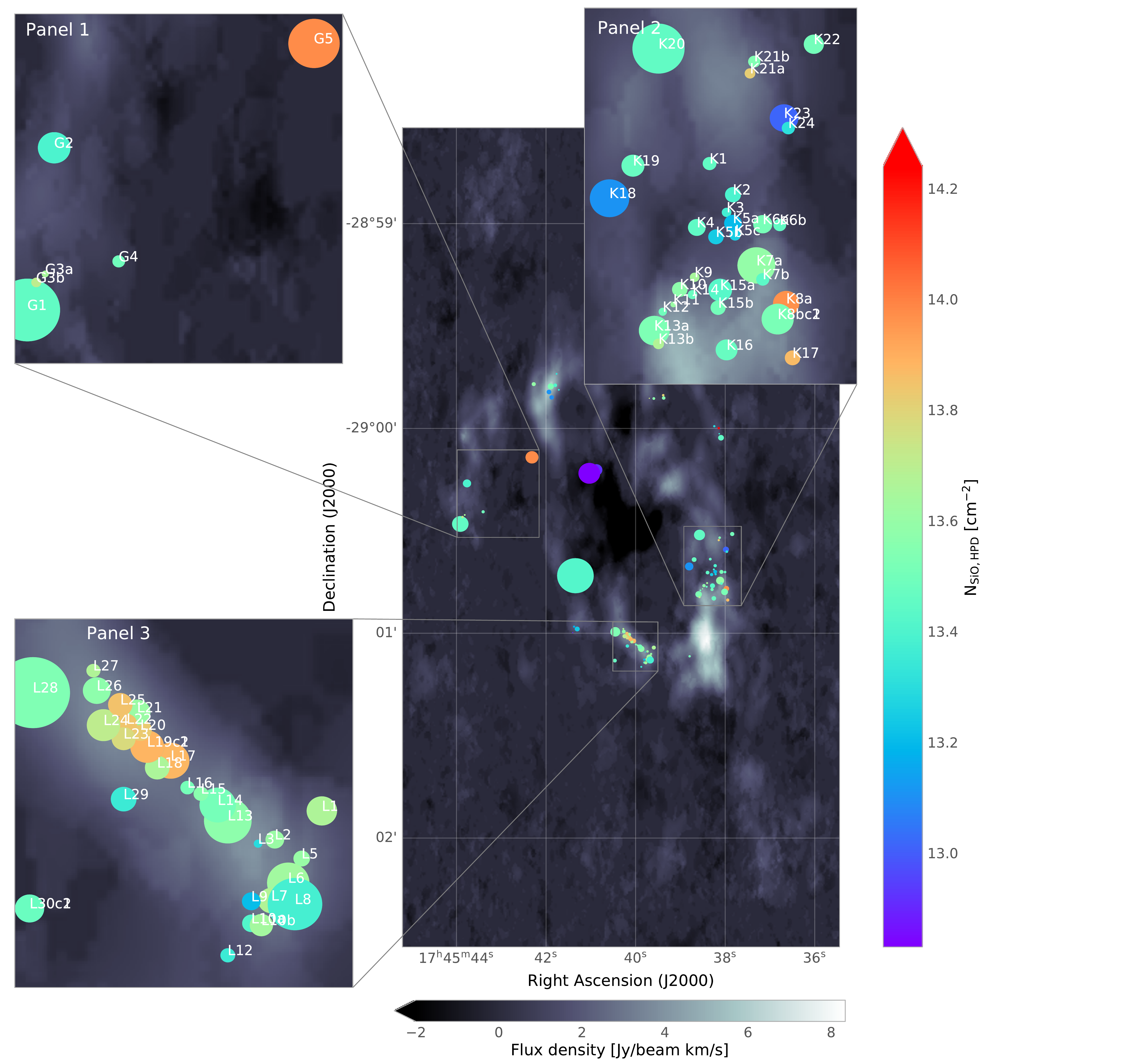}
    \caption{As per Figure \ref{fig:CND-n}, but here the colour of the circular regions represents \ce{SiO} column density N$_{\mathrm{SiO}}$ towards each observation outlined in Section \ref{sec:obs}.}
    \label{fig:CND-N-sio}
\end{figure*}

Much like Figure \ref{fig:CND-n}, the \ce{SiO} column density in Figure \ref{fig:CND-N-sio} shows little variance across the dataset. Figure \ref{fig:CND-N-sio} shows just over an order of magnitude's variation. The effects of temperature and density also appear to be negligible in this case as the regional and local variation seen in Figures \ref{fig:CND-T} and \ref{fig:CND-n} are not observed, or indeed reflected, here. For example, Panel 1 in Figure \ref{fig:CND-N-sio} shows almost uniform N$_{\mathrm{SiO}}$ in the range $10^{13.4} > N_{\mathrm{SiO}} < 10^{13.8}$ \si{\per\centi\meter\squared} despite the temperature and density varying widely. A noticeable feature of Panel 1 is the additional uniformity in N$_{\mathrm{SiO}}$ for sources within the \ce{HCN} filament. Sources outside of this filament, e.g. G5, are found to exhibit a different, larger column density.

Notable features within Panel 2 show that a tentative trend indicating the brighter the underlying \ce{HCN} emission then the larger the \ce{SiO} column density. For example, the upper half of the panel, where \ce{HCN} emission is less intense, showing marginally lower column densities than the lower half where \ce{HCN} emission is more intense. Curiously, K5a, K5b and K5c are observed to have lower column density of N$_{\mathrm{SiO}} \approx 10^{13.3}$ \si{\per\centi\meter\squared} than their surrounding observations. This could indicate that the surrounding gas acts to shield the K5 sources and thereby inhibit the emission of \ce{SiO}. This is further confirmed when noting that the gas density of K5a, K5b and K5c are consistent with their surroundings. The lower half of Panel 2 is much more uniform of N$_{\mathrm{SiO}} \approx 10^{13.6}$ \si{\per\centi\meter\squared} with the exception of K8a which has column density N$_{\mathrm{SiO}} \approx 10^{14.0}$ \si{\per\centi\meter\squared}. The inner CND ridge (K9 to K13b) all show uniform and consistent column density of N$_{\mathrm{SiO}} \approx 10^{13.6}$ \si{\per\centi\meter\squared} with this agreement extending behind the ridge until K8a. Uniquely, the previously identified ridge indicated by K12, K13a and K13b shows no distinct \ce{SiO} behaviour relative to its surroundings, thus indicating that any interaction between the Galactic Centre Mini Spiral and the CND towards this region is not having an effect on the \ce{SiO} chemistry.

Panel 3 is not dissimilar in broad trend to Panel 2 in that it shows the same agreement trends. The larger N$_{\mathrm{SiO}}$ is located towards the observation's Northernmost tip, with the column density decreasing towards the South of this complex. Interestingly, this region of relatively large N$_{\mathrm{SiO}}$ sits within an \ce{HCN} dark region. By application, \ce{HCN} traces dense molecular gas and thus shows that, at least in this region, that regions of large \ce{SiO} column density are found in dense regions of gas.

Directly North of Panel 2 is another grouping of observations that houses the largest derived value of N$_{\mathrm{SiO}}$ at $\approx 10^{14.2}$ \si{\per\centi\meter\squared}. The surrounding observations are also observed to have N$_{\mathrm{SiO}}$ towards the upper-half of the derived scale, though these observations are consistent with the values of N$_{\mathrm{SiO}}$ derived towards the series of observations in Panels 2 and 3. This is again consistent with the underlying \ce{HCN} emission tracing dense regions within the CND, ergo grain-processing as described in Section \ref{sec:intro}.

\begin{figure*}
    \centering
    \includegraphics[scale=0.47]{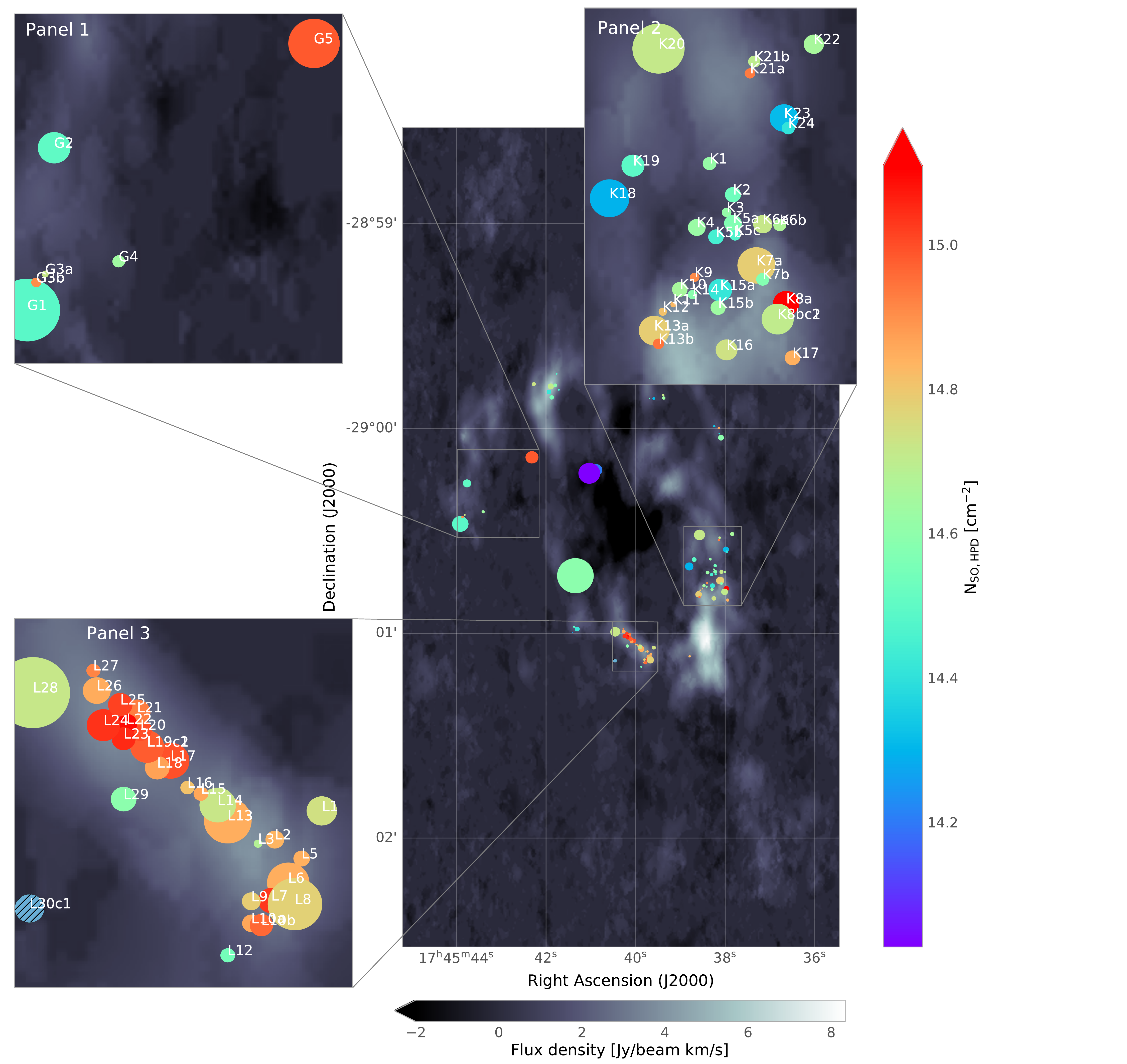}
    \caption{As per Figure \ref{fig:CND-N-sio}, but here the colour of the circular regions represents \ce{SO} column density N$_{\mathrm{SO}}$ towards each observation outlined in Section \ref{sec:obs}.}
    \label{fig:CND-N-so}
\end{figure*}

Figure \ref{fig:CND-N-so} shows that almost all sources can be considered to have N$_{\mathrm{SO}} > 10^{14}$ \si{\per\centi\meter\squared}. A grouping of $3$ observations immediately south of Sgr A* houses the lowest value of N$_{\mathrm{SO}}$ within our data at $\approx 10^{14}$ \si{\per\centi\meter\squared}.

Panel 1 echoes a similar trend to its \ce{SiO} column densities, with the largest column density in this Panel - G5 - residing outside of the main \ce{HCN} filament. Interestingly, G3b is here shown to be N$_{\mathrm{SO}} \approx 10^{15}$ \si{\per\centi\meter\squared}, larger than it's surroundings in G1 and G2 that show consistent column densities of N$_{\mathrm{SO}} > 10^{14.5}$ \si{\per\centi\meter\squared}. G3a and G4 are also consistent, with N$_{\mathrm{SO}} > 10^{14.7}$ \si{\per\centi\meter\squared}.

Panel 2 again shows a number of interesting trends, specifically a degree of non-uniformity throughout the derived values in the main grouping. We observe that whilst the local agreement between neighbouring observations is again evident here, it is more evident in the upper half of the panel than in the lower half. For instance, the collection of sources from K1 to K6b all show N$_{\mathrm{SO}} \approx 10^{14.8}$ \si{\per\centi\meter\squared}. To the South of this complex we find that the observations along the inner boundary ridge (K9 to K13b) of the CND show large column densities towards K9 and K13b, with the column density in general decreasing towards the centre of this line towards K11. K8a also stands out as a high column density source, especially in relation to its lesser column density neighbours. However, the derived column densities for these sources span around $1$ order of magnitude, making the individual variation from observation to observation slight. This is also true when applied to the interaction ridge, which is here observed to be less uniform than in \ce{SiO}. However, this behaviour is consistent with the $\mathrm{I_{\ce{SiO}}/I_{\ce{SO}}}$ ratio being consistent across the ridge.

Panel 3 again exhibits a similar trend to that seen in Figure \ref{fig:CND-N-sio}, though there is even less variation in N$_{\mathrm{SO}}$ throughout this region. The \ce{HCN} dark region towards the Northernmost tip of this grouping may again be considered to house the largest values of N$_{\mathrm{SO}} \approx 10^{15}$ \si{\per\centi\meter\squared}.

The dense N$_{\mathrm{SiO}}$ observation to the North of Panel 2 in Figure \ref{fig:CND-N-sio} now shows almost unanimous agreement in N$_{\mathrm{SO}}$ with its neighbours indicating that whatever mechanism is driving its excess \ce{SiO} chemistry - relative to its surroundings - is not reflected in its \ce{SO} chemistry. 

When considering the trends in both Figures \ref{fig:CND-N-sio} and \ref{fig:CND-N-so}, we see variation in both N$_{\mathrm{SiO}}$ and N$_{\mathrm{SO}}$ throughout the data, though this variation is slight in both dimensions and likely not large enough to indicate any localised energetic events acting as vectors for excess emission in the vast majority of cases. Importantly, the derived \ce{SiO} and \ce{SO} column densities are larger than the expected quiescent column densities. The CND wide nature of this enhancement suggests that the mechanism driving this excess is universal, and likely grain-processing and/or desorption.

\section{Formation mechanisms} \label{sec:emission}

It is clear that all sources in our data show enhancement of \ce{SO} and \ce{SiO}. As already discussed, each of these molecules is a known tracer of grain-processing activity, owing to both \ce{Si} and \ce{S} being far more abundant on the grain surfaces than they are in the gas-phase. Indeed, \ce{SiO} and \ce{SO} do not form readily in the gas-phase without \ce{Si} or \ce{S} trapped in/on the grains being released in to the gas phase for subsequent reaction with \ce{O} e.g. \ce{Si + O -> SiO}. Conversely, \ce{CH_{3}OH} is thought to form readily on grain surfaces via successive hydrogenation of \ce{CO}, where it is subsequently desorbed in to the gas-phase by an energetic mechanism such as shock activity (e.g. \citet{jShocks}) or photodesorption (e.g. \citet{ch3ohPhotoDesorption}).

Previous studies of the Galactic Center on large scales such as those by \citet{galCenDustTemp} and \citet{galacticCenterCosmicRays} have shown that the gas and dust are uncoupled and that the dust temperature $\mathrm{T_{dust}} \leq 40$ \si{\kelvin}, well below the temperature at which thermal desorption becomes significant. When considering smaller scales such as the CND, \citet{galCenKuiper} theorise that the dust temperature varies positionally from 40 \si{\kelvin} to 200 \si{\kelvin}, with the variation likely a result of heating by the OB star population. Furthermore, our analysis has also shown that the density of the most dense clumps to be $\mathrm{n_{H}} \approx 10^{7}$ \si{\per\centi\meter\cubed}, a density at which the gas and dust are likely coupled, either fully or partially \citep{depletionThermalCoupling}. As a result, thermal desorption could also be dominant within sub-regions of the CND. This naturally poses the question: what physical event(s) are dominant in driving this grain-processing, and are these events associated with specific events such as star formation, or is it instead energetic activity associated with Sgr A* and its surroundings?  

In the context of star formation, driven shock action is an efficient and ubiquitous processor of grains, so much so that molecules like \ce{SiO} and \ce{SO} are often used exclusively as tracers of shocks in star forming regions \citep{SiOShockTracer, sulphurShockTracer}. As already noted, the OB star population in the Central Cavity could raise the photoionisation rate towards the inner edge of the disc, making photodesorption of grain mantle species an efficient desorption mechanism here. Dense regions further in to the disc may also be subject to photodesorption via secondary photons. Crucially, such widespread desorption activity could act to pre-process the grain mantles, rendering molecules such as \ce{SiO} and \ce{SO} unreliable shock tracers in the CND. 

Furthermore, the uniform enhancement of \ce{SiO} and \ce{SO} observed throughout our observations is indicative of a large scale process driving this desorption. Star formation would be expected to occur in a structured manner, to the extent that its signatures would be concentrated to relatively small regions of the CND. These signatures are dependent on the evolutionary state of the object; they could manifest as cold and dense self-gravitating regions, or hot and dense gas swept up in outflows. 

K13a is, however, one of only 2 observations in our sample that shows significant \ce{OCS} emission. K13a's surroundings do not exhibit similar \ce{OCS} lines, indicating highly localised phenomena to the K13a locale. The only other \ce{OCS} positive source is G2 which, unlike K13a, exhibits \ce{CH_{3}OH} and \ce{H_{2}CS} emission. Within K13a we note that the measured linewidth of the \ce{OCS} line is a factor of 2 less than the measured linewidths of the \ce{SiO} and \ce{SO} lines. This is in spite of all lines towards K13a having consistent measured velocities. Such behaviour could indicate that the \ce{SiO} and \ce{SO} lines are tracing a different thermal component to the \ce{OCS} line, hence we disregard the \ce{OCS} line towards the position when fitting for K13a. As a result, the assumption of LTE should be cautioned towards this region. 

Interestingly, when considering the G-pointing we find that G2 does not exhibit similar behaviour, though there is minor variation in measured linewidth across the detected lines. \ce{CH_{3}OH} and \ce{H_{2}CS} emission are widespread throughout the G pointing, creating a unique subset of observations within our sample. Unfortunately, the temperature towards K13a remains unconstrained. However, if we consider the point of highest posterior density towards K13a as absolute then $\mathrm{T_{K13a}} \approx 500$ \si{\kelvin}. G2 has temperature $\mathrm{T_{G2}} \approx 600$ \si{\kelvin}. As such, the presence of \ce{OCS} could indicate high-temperature chemistry occurring within these regions. The non-detections of \ce{CH_{3}OH} and \ce{H_{2}CS} in K13a is puzzling, though K13a is known to sit along the Galactic Center Mini Spiral - CND ridge and thus could be indicative of locally different chemistry. Conversely, \ce{OCS} tracing a separate thermal component towards K13a, and the absence of this towards G2, potentially alludes to this interaction. Nevertheless, both of these regions require further chemical follow up study to confirm. 

One may broadly estimate whether the derived conditions towards the positions in this study are indicative of star formation by considering density, mass and pressure estimates. Considering G2 we note that the Roche Density \citep{rocheDensity}, assuming an approximate distance from Sgr A* of 3.0 pc, is $\approx 6 \times 10^{6}$ \si{\per\centi\meter\cubed}, which is marginally smaller than the derived gas density towards G2 thus implying tidal stability. Considering an estimate of the mass, assuming that G2 is spherical and uniform with average density $n_{\mathrm{H}} \approx 10^{6}$ \si{\per\centi\meter\cubed} and radius equivalent to $\sim$ 3 beams we find its mass $M_{\mathrm{uniform}} \approx 22 \times 10^{3} $ $M_{\odot}$. For G2 to be gravitationally bound with linewidth $\Delta v \approx 20$ \si{\kilo\meter\per\second} then its mass would need to exceed the virial mass $M_{\mathrm{vir}}$. Using the same process as \citet{Christopher2005}, we find that $M_{\mathrm{vir}}$ towards G2 is $M_{\mathrm{vir}} \approx 40 \times 10^{3}$ $M_{\odot}$, only marginally larger than $M_{\mathrm{uniform}}$. As a result, G2 could very well be gravitationally bound within the limits of the assumptions applied. Additionally, another potentially bound source is L28 within the L pointing. L28 has uniform mass $M_{\mathrm{uniform}} \approx 40 \times 10^{3} $ $M_{\odot}$ and virial mass $M_{\mathrm{vir}} \approx 48 \times 10^{3}$ $M_{\odot}$, and shows the closest agreement between uniform and virial masses within our sample. For G5, we note $M_{\mathrm{uniform}} \approx 1 \times 10^{3} $ $M_{\odot}$ and virial mass $M_{\mathrm{vir}} \approx 4 \times 10^{3}$ $M_{\odot}$. However, the recovered gas density for G5 of $\mathrm{n_{H}} \approx 10^{7}$ \si{\per\centi\meter\cubed} is therefore larger than the equivalent Roche Density at 3.0 pc, implying that G5 could be able to overcome the tidal barrier to bind itself gravitationally. This assumes that G5 is one discrete, monolithic clump. Given that G5's physical conditions render it such an outlier amongst our observed regions, it is possible that this region is actually multiple clumps along the line of sight whose spectra sum to produce a single spectra, in turn creating the illusion of a spectra consistent with one unified clump.

We apply a similar process in order to estimate the total molecular gas mass contained within the CND. This can be expressed as $M_{\mathrm{CND}} \approx \pi (R_{\mathrm{out}}^{2} - R_{\mathrm{in}}^{2}) h \rho$ where $h$ represents the height of the CND. Based on the underlying \ce{HCN} maps in Figure \ref{fig:sio-so-ratio} we  assume that this is, on average, $\approx 0.5$ pc. For comparison, we utilise the same CND radii reported in \citet{tsuboiCNR}: $R_{in} \approx 1.5$ pc and $R_{out} \approx 2.0$ pc. Additionally, we assume that the average CND density is equivalent to the modal density determined previously: $\approx 5 \times 10^{5}$ \si{\per\centi\meter\cubed}. This yields a total mass of $M_{\mathrm{CND}} \approx 4 \times 10^{4}$ $M_{\odot}$. \citet{tsuboiCNR} estimate the molecular gas mass of the CND based on LTE assumptions to be $3 \times 10^{4}$ $M_{\odot}$, marginally smaller than our estimate. However, we caution that our observations do not sample the full extent of the CND, especially towards \ce{HCN} faint regions such as those to the North of Sgr A*. In this instance, the use of the modal density derived from our observations - which deliberately target dense gas - could result in us overestimating the mean density of the CND, in turn biasing our mass estimate. 

When considering both thermal and turbulent pressure, one may express the thermal pressure as $p_{\mathrm{therm}} / k_{B} \approx n T$ and the turbulent pressure as $p_{\mathrm{turb}} / k_{B} \approx \rho \Delta v^{2}/k_{B}$. For $\mathrm{T} \approx 600$ \si{\kelvin}, $n_{\mathrm{H}} \approx 10^{6}$ \si{\per\centi\meter\cubed} and $\Delta v \approx 20$ \si{\kilo\meter\per\second} then $p_{\mathrm{therm}} / k_{B} \approx 10^{9}$ \si{\per\centi\meter\cubed\kelvin} and $p_{\mathrm{turb}} / k_{B} \approx 10^{11}$ \si{\per\centi\meter\cubed\kelvin}. Therefore, for G2 the turbulent pressure dominates over the thermal pressure in spite of the high estimated temperature towards this region. In fact, all positions within the G, K and L pointings are dominated by turbulence, despite their wide range of estimated temperatures. Additionally, this is seemingly independent of their potentially bound states, though a more rigorous determination of mass based on LTE assumptions and chemical abundances is required to identify more accurate clump masses to compare to their virial counterparts. 

% thus indicating that in spite of the exceedingly large temperature it is in fact ram pressure or turbulence that is dominating the gas in this region. Consequently, K13a is unlikely to be gravitationally bound.

% Nevertheless, to identify the physical cause of the observed complexity one must defer to chemical models that are informed by this study. Such an approach is the only method of identifying the underlying causality within the CND, in addition to determining accurate estimates of quantities required to derive more accurate mass estimates, especially in relation to the curious sources of G2 and K13a. 

\section{Conclusions} \label{sec:conclude}

In this paper, we have presented sub-arcsecond observations from 8 ALMA Band 7 (272 -- 375 \si{\giga\hertz}) pointings towards the Circumnuclear disk (CND) of Sgr A*. We detect clear and distinct \ce{SiO (8_{7} - 7_{6})} and \ce{SO (7 - 6)} emission towards 98 positions within these pointings. Additionally, we observe widespread \ce{CH_{3}OH (2_{1,1} - 2_{0,2})} emission as well as more localised emission of \ce{H_{2}CS (9_{1,9} - 8_{1,8})} and \ce{OCS (25 - 24)} towards a subset of these 98 positions. 

We have used a Bayesian Inference routine coupled to radiative transfer to infer the physical conditions towards each observation based on the detected line fluxes. Based on these results, we find the following:

\begin{itemize}
    \item The temperature of the regions probed by the observations is in general $\mathrm{T} < 500$ \si{\kelvin}, approximately consistent with that derived previously towards similar clumps. Most of the estimated densities within this work are in good agreement with the previously reported values towards clumps within the CND at $\mathrm{n_{H}} \lessapprox 10^{6}$ \si{\per\centi\meter\cubed}.
    
    \item The ubiquity of \ce{SiO} and \ce{SO} detections, including their large derived column densities, is consistent with large scale grain-processing occurring within the CND. However, we find that multiple desorption mechanisms, both thermal and non-thermal (e.g. UV or cosmic-ray photodesorption), are plausible engines with which to drive grain-processing in the CND. Further chemical modelling is required to identify which processes are efficient towards the CND.
    
    \item The G2 source within the Northeastern Arm is identified as being hot ($\mathrm{T_{kin}} \approx 600$ \si{\kelvin}) and dense ($\mathrm{n_{H}} \approx 10^{6}$ \si{\per\centi\meter\cubed}). \ce{SiO (8_{7} - 7_{6})} and \ce{SO (7 - 6)} emission is identified towards it, as well as \ce{CH_{3}OH (2_{1,1} - 2_{0,2})}, \ce{H_{2}CS (9_{1,9} - 8_{1,8})} and \ce{OCS (25 - 24)} lines. Basic calculations indicate that this region is potentially gravitationally bound and is turbulent pressure dominated. It is therefore a source of immense interest for follow up observations and study. 
    
    \item L28 is identified as having mass almost exactly consistent with that required for binding to occur, though turbulence dominates here as well. L28 does not show any outstanding feature in terms of its line detections or its recovered physical quantities, rendering it somewhat of an outlier amongst the potentially bound sources identified. Nevertheless, more accurate mass estimates in follow up studies based on complex chemistry can help constrain whether this region is indeed bound or not.
    
    \item Similarly, G5 is found to have the largest recovered density within our data of $\mathrm{n_{H}} \approx 10^{7}$ \si{\per\centi\meter\cubed} whilst also exhibiting one of the lowest recovered temperatures at $\mathrm{T_{kin}} \approx 60$ \si{\kelvin}. G5's high density is sufficient to overcome tidal effects prohibiting the object from being gravitational bound. Basic mass estimates also support this finding.
    
    \item Another position of chemical interest is K13a. K13a shows detections of \ce{SiO (8_{7} - 7_{6})} and \ce{SO (7 - 6)} as well as \ce{OCS (25 - 24)} - the only other position other than G2 to exhibit the \ce{OCS} line. K13a is located along a ridge of the CND that is interacting with the Galactic Centre Mini Spiral. Further follow up study is required to understand the role that the ionised gas within the Galactic Centre Mini Spiral plays in the chemistry of this ridge.
    
    \item Using a similar mass derivation approach, we find that the total molecular gas mass contained within the CND between 1.5 - 2.0 pc is $M_{\mathrm{CND}} \approx 4 \times 10^{4}$ $M_{\odot}$.
    
\end{itemize}

\section*{Acknowledgments}
    TAJ is funded by an STFC studentship, and thanks the STFC accordingly. SV acknowledges the European Research Council (ERC) Advanced Grant MOPPEX 833460 as well as the European Union's Horizon 2020 research and innovation program under the Marie Skodowska-Curie grant agreement No. 811312 for the project ``Astro-Chemical Origins'' (ACO). We thank J. Holdship for their discussion, opinions and associated improvements on aspects of this work. We also thank the anonymous referee and editors for their comments, suggestions and improvements that have enriched and clarified this work. 
    This paper makes use of the following ALMA data: ADS/JAO.ALMA\#2013.1.01242.S. ALMA is a partnership of ESO (representing its member states), NSF (USA) and NINS (Japan), together with NRC (Canada), MOST and ASIAA (Taiwan), and KASI (Republic of Korea), in cooperation with the Republic of Chile. The Joint ALMA Observatory is operated by ESO, AUI/NRAO and NAOJ.

\software{emcee \citep{emcee}, ChainConsumer \citep{chainConsumer}, RADEX \citep{radex}, CASA \citep[4.7.0;][]{casa}}


\bibliography{references}
\bibliographystyle{aasjournal}

\appendix 
\label{sec:appendix}

\section{Individual spectra} \label{sec:spectra}

\figsetstart
\figsetnum{9}
\figsettitle{Individual spectra}

\figsetgrpstart
\figsetgrpnum{9.1}
\figsetgrptitle{Spectra for position D1 }
\figsetplot{src_d1.pdf}
\figsetgrpnote{An example individual spectra. Spectra for each position in which at least one line was identified are available online.}
\figsetgrpend

\figsetgrpstart
\figsetgrpnum{9.2}
\figsetgrptitle{Spectra for position D2 }
\figsetplot{src_d2.pdf}
\figsetgrpnote{An example individual spectra. Spectra for each position in which at least one line was identified are available online.}
\figsetgrpend

\figsetgrpstart
\figsetgrpnum{9.3}
\figsetgrptitle{Spectra for position D3 }
\figsetplot{src_d3.pdf}
\figsetgrpnote{An example individual spectra. Spectra for each position in which at least one line was identified are available online.}
\figsetgrpend

\figsetgrpstart
\figsetgrpnum{9.4}
\figsetgrptitle{Spectra for position D4 }
\figsetplot{src_d4.pdf}
\figsetgrpnote{An example individual spectra. Spectra for each position in which at least one line was identified are available online.}
\figsetgrpend

\figsetgrpstart
\figsetgrpnum{9.5}
\figsetgrptitle{Spectra for position D5 }
\figsetplot{src_d5.pdf}
\figsetgrpnote{An example individual spectra. Spectra for each position in which at least one line was identified are available online.}
\figsetgrpend

\figsetgrpstart
\figsetgrpnum{9.6}
\figsetgrptitle{Spectra for position D6 }
\figsetplot{src_d6.pdf}
\figsetgrpnote{An example individual spectra. Spectra for each position in which at least one line was identified are available online.}
\figsetgrpend

\figsetgrpstart
\figsetgrpnum{9.7}
\figsetgrptitle{Spectra for position D7 }
\figsetplot{src_d7.pdf}
\figsetgrpnote{An example individual spectra. Spectra for each position in which at least one line was identified are available online.}
\figsetgrpend

\figsetgrpstart
\figsetgrpnum{9.8}
\figsetgrptitle{Spectra for position G1 }
\figsetplot{src_G1.pdf}
\figsetgrpnote{An example individual spectra. Spectra for each position in which at least one line was identified are available online.}
\figsetgrpend

\figsetgrpstart
\figsetgrpnum{9.9}
\figsetgrptitle{Spectra for position G2 }
\figsetplot{src_G2.pdf}
\figsetgrpnote{An example individual spectra. Spectra for each position in which at least one line was identified are available online.}
\figsetgrpend

\figsetgrpstart
\figsetgrpnum{9.10}
\figsetgrptitle{Spectra for position G3a }
\figsetplot{src_G3a.pdf}
\figsetgrpnote{An example individual spectra. Spectra for each position in which at least one line was identified are available online.}
\figsetgrpend

\figsetgrpstart
\figsetgrpnum{9.11}
\figsetgrptitle{Spectra for position G3b }
\figsetplot{src_G3b.pdf}
\figsetgrpnote{An example individual spectra. Spectra for each position in which at least one line was identified are available online.}
\figsetgrpend

\figsetgrpstart
\figsetgrpnum{9.12}
\figsetgrptitle{Spectra for position G4 }
\figsetplot{src_G4.pdf}
\figsetgrpnote{An example individual spectra. Spectra for each position in which at least one line was identified are available online.}
\figsetgrpend

\figsetgrpstart
\figsetgrpnum{9.13}
\figsetgrptitle{Spectra for position G5 }
\figsetplot{src_G5.pdf}
\figsetgrpnote{An example individual spectra. Spectra for each position in which at least one line was identified are available online.}
\figsetgrpend

\figsetgrpstart
\figsetgrpnum{9.14}
\figsetgrptitle{Spectra for position H1A }
\figsetplot{src_h1a.pdf}
\figsetgrpnote{An example individual spectra. Spectra for each position in which at least one line was identified are available online.}
\figsetgrpend

\figsetgrpstart
\figsetgrpnum{9.15}
\figsetgrptitle{Spectra for position H1B }
\figsetplot{src_h1b.pdf}
\figsetgrpnote{An example individual spectra. Spectra for each position in which at least one line was identified are available online.}
\figsetgrpend

\figsetgrpstart
\figsetgrpnum{9.16}
\figsetgrptitle{Spectra for position H2 }
\figsetplot{src_h2.pdf}
\figsetgrpnote{An example individual spectra. Spectra for each position in which at least one line was identified are available online.}
\figsetgrpend

\figsetgrpstart
\figsetgrpnum{9.17}
\figsetgrptitle{Spectra for position H3 }
\figsetplot{src_h3.pdf}
\figsetgrpnote{An example individual spectra. Spectra for each position in which at least one line was identified are available online.}
\figsetgrpend

\figsetgrpstart
\figsetgrpnum{9.18}
\figsetgrptitle{Spectra for position H4 }
\figsetplot{src_h4.pdf}
\figsetgrpnote{An example individual spectra. Spectra for each position in which at least one line was identified are available online.}
\figsetgrpend

\figsetgrpstart
\figsetgrpnum{9.19}
\figsetgrptitle{Spectra for position IRS7 }
\figsetplot{src_irs7.pdf}
\figsetgrpnote{An example individual spectra. Spectra for each position in which at least one line was identified are available online.}
\figsetgrpend

\figsetgrpstart
\figsetgrpnum{9.20}
\figsetgrptitle{Spectra for position J1 }
\figsetplot{src_j1.pdf}
\figsetgrpnote{An example individual spectra. Spectra for each position in which at least one line was identified are available online.}
\figsetgrpend

\figsetgrpstart
\figsetgrpnum{9.21}
\figsetgrptitle{Spectra for position J2 }
\figsetplot{src_j2.pdf}
\figsetgrpnote{An example individual spectra. Spectra for each position in which at least one line was identified are available online.}
\figsetgrpend

\figsetgrpstart
\figsetgrpnum{9.22}
\figsetgrptitle{Spectra for position J3 }
\figsetplot{src_j3.pdf}
\figsetgrpnote{An example individual spectra. Spectra for each position in which at least one line was identified are available online.}
\figsetgrpend

\figsetgrpstart
\figsetgrpnum{9.23}
\figsetgrptitle{Spectra for position J4 }
\figsetplot{src_j4.pdf}
\figsetgrpnote{An example individual spectra. Spectra for each position in which at least one line was identified are available online.}
\figsetgrpend

\figsetgrpstart
\figsetgrpnum{9.24}
\figsetgrptitle{Spectra for position K1 }
\figsetplot{src_k1.pdf}
\figsetgrpnote{An example individual spectra. Spectra for each position in which at least one line was identified are available online.}
\figsetgrpend

\figsetgrpstart
\figsetgrpnum{9.37}
\figsetgrptitle{Spectra for position K2 }
\figsetplot{src_k2.pdf}
\figsetgrpnote{An example individual spectra. Spectra for each position in which at least one line was identified are available online.}
\figsetgrpend

\figsetgrpstart
\figsetgrpnum{9.38}
\figsetgrptitle{Spectra for position K20 }
\figsetplot{src_k20.pdf}
\figsetgrpnote{An example individual spectra. Spectra for each position in which at least one line was identified are available online.}
\figsetgrpend

\figsetgrpstart
\figsetgrpnum{9.39}
\figsetgrptitle{Spectra for position K21A }
\figsetplot{src_k21a.pdf}
\figsetgrpnote{An example individual spectra. Spectra for each position in which at least one line was identified are available online.}
\figsetgrpend

\figsetgrpstart
\figsetgrpnum{9.40}
\figsetgrptitle{Spectra for position K21B }
\figsetplot{src_k21b.pdf}
\figsetgrpnote{An example individual spectra. Spectra for each position in which at least one line was identified are available online.}
\figsetgrpend

\figsetgrpstart
\figsetgrpnum{9.41}
\figsetgrptitle{Spectra for position K22 }
\figsetplot{src_k22.pdf}
\figsetgrpnote{An example individual spectra. Spectra for each position in which at least one line was identified are available online.}
\figsetgrpend

\figsetgrpstart
\figsetgrpnum{9.42}
\figsetgrptitle{Spectra for position K23 }
\figsetplot{src_k23.pdf}
\figsetgrpnote{An example individual spectra. Spectra for each position in which at least one line was identified are available online.}
\figsetgrpend

\figsetgrpstart
\figsetgrpnum{9.43}
\figsetgrptitle{Spectra for position K24 }
\figsetplot{src_k24.pdf}
\figsetgrpnote{An example individual spectra. Spectra for each position in which at least one line was identified are available online.}
\figsetgrpend

\figsetgrpstart
\figsetgrpnum{9.44}
\figsetgrptitle{Spectra for position K3 }
\figsetplot{src_k3.pdf}
\figsetgrpnote{An example individual spectra. Spectra for each position in which at least one line was identified are available online.}
\figsetgrpend

\figsetgrpstart
\figsetgrpnum{9.45}
\figsetgrptitle{Spectra for position K4 }
\figsetplot{src_k4.pdf}
\figsetgrpnote{An example individual spectra. Spectra for each position in which at least one line was identified are available online.}
\figsetgrpend

\figsetgrpstart
\figsetgrpnum{9.46}
\figsetgrptitle{Spectra for position K5A }
\figsetplot{src_k5a.pdf}
\figsetgrpnote{An example individual spectra. Spectra for each position in which at least one line was identified are available online.}
\figsetgrpend

\figsetgrpstart
\figsetgrpnum{9.47}
\figsetgrptitle{Spectra for position K5B }
\figsetplot{src_k5b.pdf}
\figsetgrpnote{An example individual spectra. Spectra for each position in which at least one line was identified are available online.}
\figsetgrpend

\figsetgrpstart
\figsetgrpnum{9.48}
\figsetgrptitle{Spectra for position K5C }
\figsetplot{src_k5c.pdf}
\figsetgrpnote{An example individual spectra. Spectra for each position in which at least one line was identified are available online.}
\figsetgrpend

\figsetgrpstart
\figsetgrpnum{9.49}
\figsetgrptitle{Spectra for position K6A }
\figsetplot{src_k6a.pdf}
\figsetgrpnote{An example individual spectra. Spectra for each position in which at least one line was identified are available online.}
\figsetgrpend

\figsetgrpstart
\figsetgrpnum{9.50}
\figsetgrptitle{Spectra for position K6B }
\figsetplot{src_k6b.pdf}
\figsetgrpnote{An example individual spectra. Spectra for each position in which at least one line was identified are available online.}
\figsetgrpend

\figsetend

\begin{figure}
\figurenum{9}
\plotone{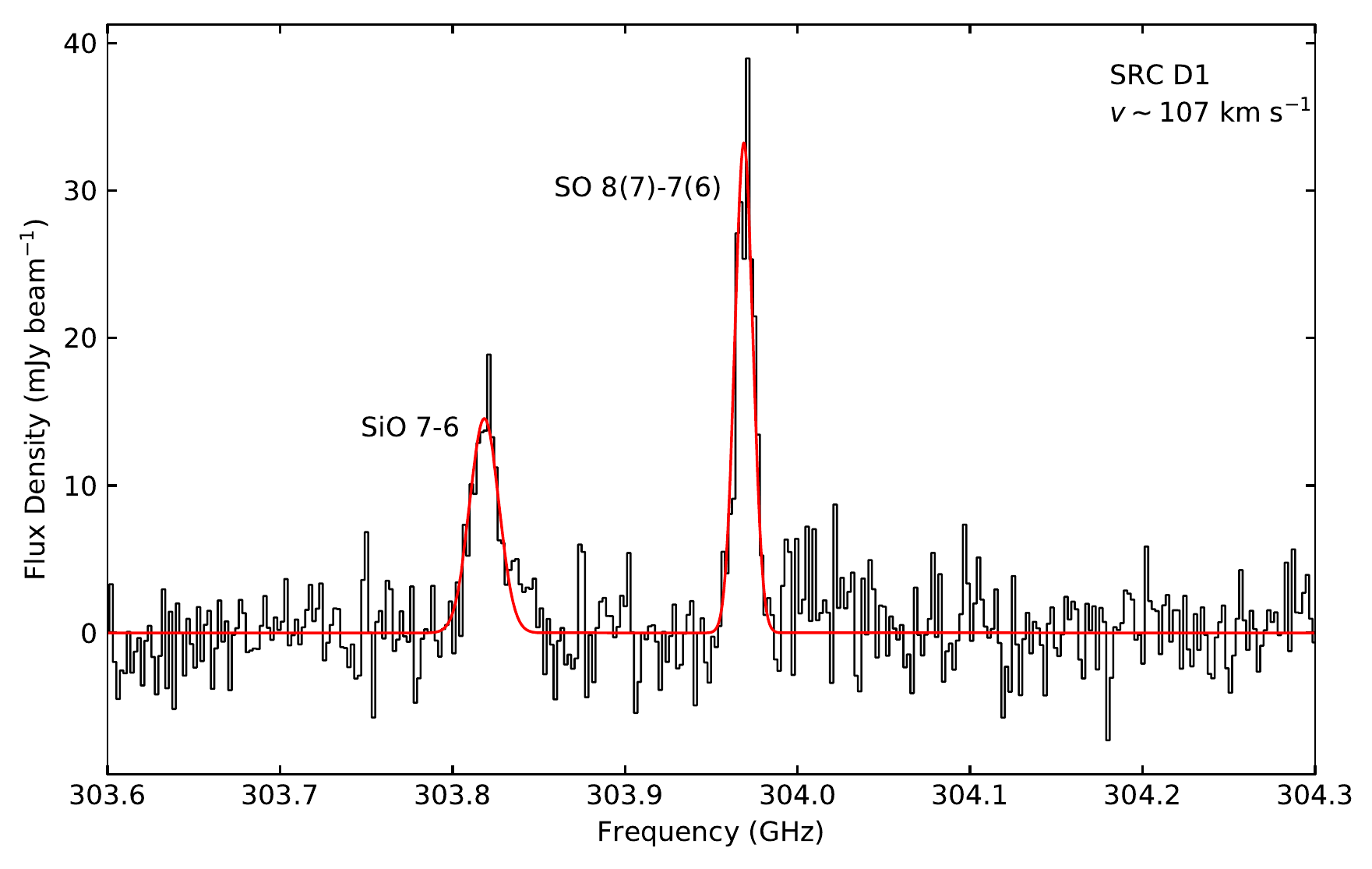}
\caption{An example individual spectra for position D1. Spectra for each position in which at least one line was identified are available in the online journal.}
\end{figure}
    
\clearpage

\section{Tabulated spectral data} \label{sec:data-table}

\begin{deluxetable}{ccccccccc}[htbp!] 
    \centering 
    \tablecaption{Tabulated summary of observations including the source designation, it's RA and Dec, extent in beam sizes, the emission line detected, its peak line flux density, the local standard of rest velocity, the linewidth and the ratio of \ce{SiO} to \ce{SO} flux densities where relevant.}
    \tablehead{
        Source & $\alpha$ & $\delta$ & Extent & Emission & Flux density & v & $\Delta v$ & $\mathrm{I}_\mathrm{SiO}/\mathrm{I}_\mathrm{SO}$ \\
        & (J2000) & (J2000) & (beam) & Line & (mJy/beam) & (km s$^{-1}$) & (km s$^{-1}$)  & 
    }
    \startdata
        N3     & $17^h45^m40.62^s$ & $-29^\circ00^\prime 24.05^{\prime \prime}$ & 1.00 &
      SiO 7--6 &  58.6 $\pm$  3.6 & -25.56 &  27.6  & \\
        IRS7   & $17^h45^m40.04^s$ & $-29^\circ00^\prime 22.76^{\prime \prime}$ & 1.00 &
      SO 8(7)--7(6) &  8.34 $\pm$  2.6 & 32.80 &  20.9 &\\
        N8     & $17^h45^m40.85^s$ & $-29^\circ00^\prime 12.88^{\prime \prime}$ & 1.00 &
      SiO 7--6 &  11.4 $\pm$  5.3 & 39.52 &  11.8 & 0.56 $\pm$ 0.28\\
        &&&&    SO 8(7)--7(6) &  20.5 $\pm$  4.1 & 38.73 &  15.0 &\\
        N7a    & $17^h45^m40.88^s$ & $-29^\circ00^\prime 12.23^{\prime \prime}$ & 10.65 &
      SiO 7--6 &  4.31 $\pm$ 0.47 &   41.23 &  13.9 & 0.56 $\pm$ 0.07\\
        &&&&    SO 8(7)--7(6) &  7.66 $\pm$  0.40 & 40.1649 & 16.2 &\\
        N7b    & $17^h45^m41.04^s$ & $-29^\circ00^\prime 13.24^{\prime \prime}$ & 19.59 &
      SiO 7--6 &   3.7 $\pm$ 0.22 & 39.33 &  13.9 & 0.67 $\pm$ 0.05\\
        &&&&    SO 8(7)--7(6) &  5.53 $\pm$  0.20 & 41.1081 &  15.5 & \\
        &&&&    CH$_3$OH 2(1,1)--2(0,2) &  2.43 $\pm$ 0.36 & 37.68 &  8.68 &\\
        C1     & $17^h45^m41.35^s$ & $-29^\circ00^\prime 43.23^{\prime \prime}$ & 33.03 &
      SiO 7--6 &  8.13 $\pm$ 0.35 & -42.7327 &  24.8 & 0.48 $\pm$ 0.03 \\
        &&&&    SO 8(7)--7(6) &  17.0 $\pm$ 0.47 & -42.5732 &  18.6 \\
        G1    & $17^h45^m43.92^s$ & $-29^\circ00^\prime 28.07^{\prime \prime}$ & 14.8 & 
        	SiO 7--6 & 5.84 $\pm$ 0.42 &  67.47 &  35.4   &0.88 $\pm$ 0.08\\
        &&&&    SO 8(7)--7(6) & 7.77 $\pm$  0.5 &   64.58  &  30.3& \\
        &&&&    CH$_3$OH 2(1,1)-2(0,2) & 4.49 $\pm$ 0.47 &   59.29 &  31.7& \\
        G2    & $17^h45^m43.77^s$ & $-29^\circ00^\prime 16.19^{\prime \prime}$ & 73.6 &
            SiO 7--6  & 9.53 $\pm$ 0.10 &   56.84 &  19.7&0.75 $\pm$ -0.01\\
        &&&&    SO 8(7)--7(6) &    16.7 $\pm$ 0.13 &   55.64 & -15.0& \\
        &&&&    CH$_3$OH 2(1,1)--2(0,2) &    9.57 $\pm$ 0.12 &   56.14 &  15.6& \\
        &&&&    H$_2$CS 9(1,9)--8(1,8) &   6.6 $\pm$ 0.19 &   54.66 &  9.99& \\
        &&&&    OCS 25--24 &1.72 $\pm$ 0.16 &   54.34 &  12.5& \\
        G3a    & $17^h45^m43.82^s$ & $-29^\circ00^\prime 25.43^{\prime \prime}$ & 1.59 &
            SiO 7--6  & 32.2 $\pm$  3.8 &   53.65 &  9.86&0.92 $\pm$ 0.15\\
        &&&&    SO 8(7)--7(6) & 39.2 $\pm$  4.3 &   52.77 &  8.82& \\
        &&&&    CH$_3$OH 2(1,1)--2(0,2) & 29.1 $\pm$  3.9 &   54.01 &   9.7& \\
        &&&&    H$_2$CS 9(1,9)--8(1,8) & 21.2 $\pm$  4.4 &   53.24 &  8.42& \\
        G3b    & $17^h45^m43.87^s$ & $-29^\circ00^\prime 26.05^{\prime \prime}$ & 2.24 &
          SiO 7--6 & 35.1 $\pm$  3.5 &   56.66 &  10.6&0.68 $\pm$ 0.08\\
        &&&&    SO 8(7)--7(6) & 48.6 $\pm$  3.3 &   56.98 &  11.2& \\
        &&&&    CH$_3$OH 2(1,1)--2(0,2) & 22.5 $\pm$  3.0 &   57.15 &  12.2& \\
        &&&&    H$_2$CS 9(1,9)--8(1,8)  & 16.4 $\pm$  3.7 &   55.51 &  10.1& \\
    \enddata
\end{deluxetable} \label{tab:data-table}

\begin{deluxetable}{ccccccccc}[htbp!]
    \centering
    \tablecaption{}
    \tablehead{
        Source & $\alpha$ & $\delta$ & Extent & Emission & Flux density & v & $\Delta v$ & $\mathrm{I}_\mathrm{SiO}/\mathrm{I}_\mathrm{SO}$ \\
        & (J2000) & (J2000) & (beam) & Line & (mJy/beam) & (km s$^{-1}$) & (km s$^{-1}$)  & 
    }
    \startdata
        G4     & $17^h45^m43.41^s$ & $-29^\circ00^\prime 24.51^{\prime \prime}$ &  2.85 &
          SiO 7--6 & 20.3 $\pm$  3.1 &   52.59 &  11.2&0.72 $\pm$ 0.13\\
        &&&&    SO 8(7)--7(6) & 25.7 $\pm$  2.8 &   53.18 &  12.3& \\
        &&&&    CH$_3$OH 2(1,1)--2(0,2) & 14.5 $\pm$  2.7 &   52.37 &  13.0& \\
        G5     & $17^h45^m42.32^s$ & $-29^\circ00^\prime 8.55^{\prime \prime}$ & 11.6 &
            SiO 7--6  &  39.0 $\pm$  4.7 &   55.18 &  15.5&0.93 $\pm$ 0.15\\
        &&&&    SO 8(7)--7(6) & 36.1 $\pm$  4.0 &   56.19 &  18.1& \\
        &&&&    CH$_3$OH 2(1,1)--2(0,2) & 39.9 $\pm$  4.9 &   54.99 &  14.9& \\
        &&&&    H$_2$CS 9(1,9)--8(1,8) &  17.6 $\pm$  5.3 &   55.48 &  13.8& \\
        D1     & $17^h45^m42.28^s$ & $-28^\circ59^\prime 47.11^{\prime \prime}$ & 3.69 &
              SiO 7--6 & 14.5 $\pm$  1.4 &   107.1 &  19.6&0.7 $\pm$ 0.08\\
        &&&&    SO 8(7)--7(6) & 33.2 $\pm$  2.2 &   107.5 &  12.2 & \\
        D2     & $17^h45^m41.90^s$ & $-28^\circ59^\prime 47.76^{\prime \prime}$ & 5.52 &
              SiO 7--6 & 14.4 $\pm$  0.7 &   106.3 &  17.3&0.65 $\pm$ 0.04\\
        &&&&    SO 8(7)--7(6) &  25.4 $\pm$  0.8 &   106.0 &  15.1& \\
        D3c1     & $17^h45^m41.94^s$ & $-28^\circ59^\prime 49.41^{\prime \prime}$ & 4.2 &
              SiO 7--6 &   4.61 $\pm$ 0.58 &   109.8 &  21.3& 0.55 $\pm$ 0.0.08\\
        &&&&          &  4.51 $\pm$  0.6 &   75.89 &  20.5& 0.46 $\pm$ 0.07\\
        &&&&    SO 8(7)--7(6) & 11.2 $\pm$ 0.77 &   108.0 &  15.8& \\
        &&&&          & 9.55 $\pm$ 0.59 &    76.10 &  21.1& \\
        D4     & $17^h45^m41.80^s$ & $-28^\circ59^\prime 47.44^{\prime \prime}$ & 3.3 &
              SiO 7--6 & 16.3 $\pm$  1.3 & 107.31 &  11.7&0.62 $\pm$ 0.06\\
        &&&&    SO 8(7)--7(6) &  23.3 $\pm$  1.2 & 106.34 &  13.2& \\
        &&&&    CH$_3$OH 2(1,1)--2(0,2) & 6.53 $\pm$  1.4 & 107.233 &  11.0 \& \\
        D5     &  $17^h45^m41.88^s$ & $-28^\circ59^\prime 51.00^{\prime \prime}$ & 4.04 &
              SiO 7--6 &  4.75 $\pm$ 0.74 & 77.92 &  18.8&0.33 $\pm$ 0.05\\
        &&&&    SO 8(7)--7(6) &  12.1 $\pm$ 0.63 & 76.81 &  22.1& \\
        D6     &  $17^h45^m41.72^s$ & $-28^\circ59^\prime 48.78^{\prime \prime}$ & 1.63 & 
              SiO 7--6 &  15.5 $\pm$  3.1 & -12.20 &  9.44&0.71 $\pm$ 0.17\\
        &&&&    SO 8(7)--7(6) &  12.9 $\pm$  1.8 & -14.01 &  15.9& \\
        D7     &  $17^h45^m41.77^s$ & $-28^\circ59^\prime 44.10^{\prime \prime}$ & 1.65 &
              SiO 7--6 &  10.9 $\pm$  2.2 & 93.44 &  13.3&0.67 $\pm$ 0.16\\
        &&&&    SO 8(7)--7(6) &  19.5 $\pm$  2.6 & 93.69 &  11.1& \\
        H1a    &  $17^h45^m39.38^s$ & $-28^\circ59^\prime 51.18^{\prime \prime}$ & 3.1 &
              SiO 7--6 &  18.4 $\pm$  1.1 & 35.10 &  14.7&0.83 $\pm$ 0.07\\
        &&&&    SO 8(7)--7(6) &  25.6 $\pm$  1.3 & 34.95 &  12.7& \\
        &&&&    CH$_3$OH 2(1,1)--2(0,2) & 4.96 $\pm$ 0.88 &  35.74 &  19.3& \\
        H1b    &  $17^h45^m39.39^s$ & $-28^\circ59^\prime 50.36^{\prime \prime}$ & 2.3 &
              SiO 7--6 &  22.3 $\pm$  1.1 & 30.53 &  21.7&1.24 $\pm$ 0.1\\
        &&&&    SO 8(7)--7(6) &  18.2 $\pm$  1.1 & 30.61 &  21.4& \\
        &&&&    CH$_3$OH 2(1,1)--2(0,2) &  6.49 $\pm$  1.3 & 28.36 &  17.2& \\
        H2     & $17^h45^m39.60^s$ & $-28^\circ59^\prime 51.33^{\prime \prime}$ & 2.6 &
              SiO 7--6 &  17.5 $\pm$  1.7 & 37.15 &  15.6&1.85 $\pm$ 0.37\\
        &&&&    SO 8(7)--7(6) &  12.6 $\pm$  2.2 & 36.92 &  11.7& \\
        H3     & $17^h45^m39.70^s$ & $-28^\circ59^\prime 51.26^{\prime \prime}$ & 1.1 &
              SiO 7--6 &  14.2 $\pm$  1.7 & 53.01 &  28.2&1.78 $\pm$ 0.43\\
        &&&&    SO 8(7)--7(6) &  10.3 $\pm$  2.2 & 48.21 &  21.9& \\
        H4     & $17^h45^m39.76^s$ & $-28^\circ59^\prime 46.03^{\prime \prime}$ & 2.11 &
              SiO 7--6 &  13.5 $\pm$ 0.97 & 73.59 &  32.9&2.04 $\pm$ 0.34\\
        &&&&    SO 8(7)--7(6) &  7.24 $\pm$  1.1 & 70.22 &  30.1& \\
        J1     &  $17^h45^m38.14^s$ & $-29^\circ00^\prime 1.65^{\prime \prime}$ & 1.3 &
              SiO 7--6 &  14.3 $\pm$  2.9 &  62.28 &  11.3&0.95 $\pm$ 0.27\\
        &&&&    SO 8(7)--7(6) &  18.5 $\pm$  3.6 & 60.90 &  9.23& \\
        &&&&    CH$_3$OH 2(1,1)--2(0,2) &  8.42 $\pm$  2.4 & 60.49 &  13.7& \\
        &&&&    H$_2$CS 9(1,9)--8(1,8) &  6.57 $\pm$  1.9 & 60.76 &  17.5& \\
    \enddata
\end{deluxetable}

\begin{deluxetable}{ccccccccc}[htbp!]
    \centering
    \tablecaption{}
    \tablehead{
        Source & $\alpha$ & $\delta$ & Extent & Emission & Flux density & v & $\Delta v$ & $\mathrm{I}_\mathrm{SiO}/\mathrm{I}_\mathrm{SO}$ \\
        & (J2000) & (J2000) & (beam) & Line & (mJy/beam) & (km s$^{-1}$) & (km s$^{-1}$)  & 
    }
    \startdata
        J2     & $17^h45^m38.15^s$ & $-29^\circ00^\prime 0.01^{\prime \prime}$ & 2.2 &
              SiO 7--6 &  46.9 $\pm$  1.0 & 67.16 &  27.5&2.26 $\pm$ 0.12\\
        &&&&    SO 8(7)--7(6) &  25.0 $\pm$  1.2 & 65.70 &  22.8& \\
        J3     & $17^h45^m38.10^s$ & $-29^\circ00^\prime 2.82^{\prime \prime}$ & 5.2 &
              SiO 7--6 &  14.6 $\pm$ 0.81 & 75.19 &  14.3&0.67 $\pm$ 0.04\\
        &&&&    SO 8(7)--7(6) &  25.5 $\pm$ 0.94 & 74.50 &  12.3& \\
        &&&&    CH$_3$OH 2(1,1)--2(0,2) &  10.2 $\pm$ 0.87 & 73.31 &  13.3& \\
        &&&&    H$_2$CS 9(1,9)--8(1,8) &  6.13 $\pm$  1.6 & 75.00 &  7.43& \\
        J4     & $17^h45^m38.25^s$ & $-28^\circ59^\prime 59.44^{\prime \prime}$ & 2.06 &
              SiO 7--6 &  10.3 $\pm$  2.2 & 68.63 &  12.7&0.88 $\pm$ 0.25\\
        &&&&    SO 8(7)--7(6) &  14.6 $\pm$  2.8 & 68.91 &  10.2& \\
        &&&&    CH$_3$OH 2(1,1)--2(0,2) &  10.8 $\pm$  4.4 & 67.94 &  6.46& \\
        K1     & $17^h45^m38.34^s$ & $-29^\circ00^\prime 38.38^{\prime \prime}$ & 2.61 &
              SiO 7--6 &  16.6 $\pm$  1.8 & -5.85 &  12.6&0.67 $\pm$ 0.09\\
        &&&&    SO 8(7)--7(6) &  21.6 $\pm$  1.6 & -6.27 &  14.4& \\
        K2     & $17^h45^m38.23^s$ & $-29^\circ00^\prime 40.30^{\prime \prime}$ & 3.03 &
              SiO 7--6 &  17.5 $\pm$  2.1 & -9.06 &  10.3&0.71 $\pm$ 0.1\\
        &&&&    SO 8(7)--7(6) &  21.4 $\pm$  1.8 & -7.60 &  11.9& \\
        K3     & $17^h45^m38.26^s$ & $-29^\circ00^\prime 41.39^{\prime \prime}$ & 1.83 &
              SiO 7--6 &  11.4 $\pm$  1.8 & 4.86 &  14.9&0.55 $\pm$ 0.1\\
        &&&&    SO 8(7)--7(6) &  18.7 $\pm$  1.7 & 2.45 &  16.5& \\
        K4     & $17^h45^m38.40^s$ & $-29^\circ00^\prime 42.32^{\prime \prime}$ & 3.35 &
              SiO 7--6 &  14.6 $\pm$  1.2 & -17.53 &  13.9&0.65 $\pm$ 0.06\\
        &&&&    SO 8(7)--7(6) &  22.7 $\pm$  1.2 & -18.35 &  13.7& \\
        K5a    & $17^h45^m38.23^s$ & $-29^\circ00^\prime 42.06^{\prime \prime}$ & 3.35 &
              SiO 7--6 &  9.71 $\pm$  1.5 & -15.30 &  11.7&0.41 $\pm$ 0.07\\
        &&&&    SO 8(7)--7(6) &  13.7 $\pm$ 0.89 & -10.31 &  20.0& \\
        K5b    & $17^h45^m38.31^s$ & $-29^\circ00^\prime 42.90^{\prime \prime}$ & 2.94 &
              SiO 7--6 &  8.55 $\pm$  1.3 & -20.32 &  14.8&0.59 $\pm$ 0.11\\
        &&&&    SO 8(7)--7(6) &  16.3 $\pm$  1.5 & -20.30 &  13.1& \\
        K5c    & $17^h45^m38.22^s$ & $-29^\circ00^\prime 42.80^{\prime \prime}$ & 2.11 &
              SiO 7--6 &  8.55 $\pm$  1.3 & -20.32 &  14.8&0.59 $\pm$ 0.11\\
        &&&&    SO 8(7)--7(6) &  16.3 $\pm$  1.5 & -20.30 &  13.1& \\
        K6a    & $17^h45^m38.09^s$ & $-29^\circ00^\prime 42.12^{\prime \prime}$ & 3.63 &
              SiO 7--6 &  11.6 $\pm$  1.1 & -1.67 &  20.7&0.62 $\pm$ 0.07\\
        &&&&    SO 8(7)--7(6) &  19.4 $\pm$  1.1 & -3.04 &  20.1& \\
        K6b    & $17^h45^m38.01^s$ & $-29^\circ00^\prime 42.17^{\prime \prime}$ & 2.44 &
              SiO 7--6 &  10.3 $\pm$  1.6 & -13.41 &  19.8&0.58 $\pm$ 0.1\\
        &&&&    SO 8(7)--7(6) &  19.0 $\pm$  1.7 & -12.71 &  18.6& \\
        K7a     & $17^h45^m38.12^s$ & $-29^\circ00^\prime 44.66^{\prime \prime}$ & 7.20 &
              SiO 7--6 &  21.4 $\pm$  1.1 & -42.07 &  13.7&0.61 $\pm$ 0.04\\
        &&&&    SO 8(7)--7(6) &  35.5 $\pm$  1.1 & -41.93 &  13.5& \\
        &&&&    CH$_3$OH 2(1,1)--2(0,2) &  4.99 $\pm$ 0.94 & -39.19 &  15.4& \\
        K7b    & $17^h45^m38.09^s$ & $-29^\circ00^\prime 45.52^{\prime \prime}$ & 2.50 &
              SiO 7--6 &  12.3 $\pm$  3.0 & -50.34 &  15.3&0.7 $\pm$ 0.21\\
        &&&&    SO 8(7)--7(6) &  18.3 $\pm$  3.2 & -45.76 &  14.7& \\
        K8a    & $17^h45^m37.98^s$ & $-29^\circ00^\prime 47.01^{\prime \prime}$ & 5.00 & 
              SiO 7--6 &  25.6 $\pm$  1.5 & -43.72 &  25.9&0.69 $\pm$ 0.05\\
        &&&&    SO 8(7)--7(6) &  41.9 $\pm$  1.7 & -44.99 &  23.0& \\
        K8b    & $17^h45^m38.02^s$ & $-29^\circ00^\prime 47.97^{\prime \prime}$ & 6.10 &
              SiO 7--6 &  23.9 $\pm$ 1.7 & -40.48 &  21.9&1.90 $\pm$ 0.33\\
        &&&&           &   13.1 $\pm$ 2.1 & -10.80 &  17.8& 0.61 $\pm$ 0.11\\
        &&&&    SO 8(7)--7(6) &  13.1 $\pm$  2.1 & -41.07 &  21.0& \\
        &&&&               &  23.7 $\pm$ 2.3 &  -9.99 &  16.2& \\
        K9    & $17^h45^m38.41^s$ & $-29^\circ00^\prime 45.39^{\prime \prime}$ & 1.83 &
              SiO 7--6 &  12.4 $\pm$  1.6 & -3.19 &  26.6&0.55 $\pm$ 0.08\\
        &&&&    SO 8(7)--7(6) &  24.6 $\pm$  1.7 & -5.23 &  24.4& \\
    \enddata
\end{deluxetable}

\clearpage

\begin{deluxetable}{ccccccccc}[htbp!]
    \centering
    \tablecaption{}
    \tablehead{
        Source & $\alpha$ & $\delta$ & Extent & Emission & Flux density & v & $\Delta v$ & $\mathrm{I}_\mathrm{SiO}/\mathrm{I}_\mathrm{SO}$ \\
        & (J2000) & (J2000) & (beam) & Line & (mJy/beam) & (km s$^{-1}$) & (km s$^{-1}$)  & 
    }
    \startdata
        K10   & $17^h45^m38.48^s$ & $-29^\circ00^\prime 46.14^{\prime \prime}$ & 2.94 &
              SiO 7--6 &  13.0 $\pm$  1.6 & -24.71 &  21.0&0.8 $\pm$ 0.13\\
        &&&&    SO 8(7)--7(6) &  18.5 $\pm$  1.8 & -23.93 &  18.4& \\
        K11   & $17^h45^m38.51^s$ & $-29^\circ00^\prime 47.07^{\prime \prime}$ & 1.13 &
              SiO 7--6 &  11.8 $\pm$  3.7 & -16.19 &  22.7&0.59 $\pm$ 0.22\\
        &&&&    SO 8(7)--7(6) &  23.8 $\pm$  4.4 & -20.11 &  19.0& \\
        K12   & $17^h45^m38.56^s$ & $-29^\circ00^\prime 47.52^{\prime \prime}$ & 1.58 &
              SiO 7--6 &  13.5 $\pm$  5.2 & -28.22 &  15.4&0.48 $\pm$ 0.2\\
        &&&&    SO 8(7)--7(6) &  29.1 $\pm$  5.3 & -28.69 &  15.0& \\
        K13a  & $17^h45^m38.60^s$ & $-29^\circ00^\prime 48.68^{\prime \prime}$ & 5.85 &
              SiO 7--6 &  16.3 $\pm$  2.7 & -92.39 &  14.5&0.50 $\pm$ 0.18\\
        &&&&    SO 8(7)--7(6) &  32.37 $\pm$  2.5 & -95.18 &  14.3& \\
        &&&&    OCS 25--24 &  11.1 $\pm$  5.5 & -92.99 &  7.15& \\
        K13b  & $17^h45^m38.58^s$ & $-29^\circ00^\prime 49.50^{\prime \prime}$ & 2.09 &
              SiO 7--6 &  18.0 $\pm$  7.4 & -83.93 &  16.4&0.51 $\pm$ 0.23\\
        &&&&    SO 8(7)--7(6) &  41.4 $\pm$  8.6 & -83.68 &  14.0& \\
        K14   & $17^h45^m38.42^s$ & $-29^\circ00^\prime 46.46^{\prime \prime}$ & 1.81 &
              SiO 7--6 &  13.9 $\pm$  3.8 & -26.31 &  14.1&0.73 $\pm$ 0.25\\
        &&&&    SO 8(7)--7(6) &  17.5 $\pm$  3.5 & -23.83 &  15.3& \\
        K15a  & $17^h45^m38.29^s$ & $-29^\circ00^\prime 46.18^{\prime \prime}$ & 4.50 &
              SiO 7--6 &  19.5 $\pm$  2.3 & -9.54 &  9.75&1.01 $\pm$ 0.17\\
        &&&&    SO 8(7)--7(6) &  17.7 $\pm$  2.1 & -8.82 &  10.6& \\
        K15b   & $17^h45^m38.30^s$ & $-29^\circ00^\prime 47.27^{\prime \prime}$ & 2.85 &
              SiO 7--6 &  15.7 $\pm$  3.3 & 2.05 &  13.7&0.71 $\pm$ 0.18\\
        &&&&    SO 8(7)--7(6) &  22.7 $\pm$  3.4 & 1.17 &  13.3& \\
        K16    & $17^h45^m38.26^s$ & $-29^\circ00^\prime 49.87^{\prime \prime}$ & 4.15 &
              SiO 7--6 &  18.3 $\pm$  5.4 & -36.22 &  10.5&0.5 $\pm$ 0.17\\
        &&&&    SO 8(7)--7(6) &  31.5 $\pm$  4.7 & -35.98 &  12.2& \\
        K17    & $17^h45^m37.95^s$ & $-29^\circ00^\prime 50.36^{\prime \prime}$ & 2.94 &
              SiO 7--6 &  29.2 $\pm$  8.3 & -36.32 &  16.2&1.05 $\pm$ 0.43\\
        &&&&    SO 8(7)--7(6) &  26.9 $\pm$  8.0 & -37.58 &  16.8& \\
        K18    & $17^h45^m38.81^s$ & $-29^\circ00^\prime 40.52^{\prime \prime}$ & 7.51 &
              SiO 7--6 &  6.05 $\pm$  1.1 & -14.92 &  14.7&0.6 $\pm$ 0.13\\
        &&&&    SO 8(7)--7(6) &  9.54 $\pm$  1.0 & -16.35 &  15.6& \\
        K19    & $17^h45^m38.70^s$ & $-29^\circ00^\prime 38.51^{\prime \prime}$ & 4.37 &
              SiO 7--6 &  7.46 $\pm$ 0.73 & -29.25 &  29.4&0.91 $\pm$ 0.12\\
        &&&&    SO 8(7)--7(6) &  8.54 $\pm$ 0.76 & -27.55 &  28.2& \\
        K20    & $17^h45^m38.58^s$ & $-29^\circ00^\prime 31.29^{\prime \prime}$ & 9.93 &
              SiO 7--6 &  10.3 $\pm$  1.3 & -2.82 &  19.7&0.53 $\pm$ 0.08\\
        &&&&    SO 8(7)--7(6) &  16.0 $\pm$  1.1 & -5.92 &  24.0& \\
        K21a   & $17^h45^m38.15^s$ & $-29^\circ00^\prime 32.82^{\prime \prime}$ & 2.03 &
              SiO 7--6 &  17.5 $\pm$  2.3 & 32.31 &  27.1&0.74 $\pm$ 0.12\\
        &&&&    SO 8(7)--7(6) &  25.1 $\pm$  2.4 & 31.28 &  25.4& \\
        K21b   & $17^h45^m38.13^s$ & $-29^\circ00^\prime 32.09^{\prime \prime}$ & 2.33 &
              SiO 7--6 &  18.4 $\pm$  4.2 & 22.53 &  13.1&0.69 $\pm$ 0.19\\
        &&&&    SO 8(7)--7(6) &  28.6 $\pm$  4.4 & 21.08 &  12.3& \\
        K22    & $17^h45^m37.85^s$ & $-29^\circ00^\prime 31.02^{\prime \prime}$ & 3.83 &
              SiO 7--6 &  13.4 $\pm$  5.7 & -8.65 &  14.5 &0.66 $\pm$ 0.34\\
        &&&&    SO 8(7)--7(6) &  18.7 $\pm$  5.3 & -9.83 &  15.7& \\
        K23    & $17^h45^m37.99^s$ & $-29^\circ00^\prime 35.57^{\prime \prime}$ & 5.45 &
              SiO 7--6 &  6.86 $\pm$  1.8 &  51.19 &  10.4&0.47 $\pm$ 0.14\\
        &&&&    SO 8(7)--7(6) &  15.0 $\pm$  1.8 & 51.66 &  10.1& \\
        K24    & $17^h45^m37.97^s$ & $-29^\circ00^\prime 36.19^{\prime \prime}$ & 2.49 &
                      SiO 7--6 &  9.11 $\pm$  2.2 & -17.10 &  15.2&0.79 $\pm$ 0.24\\
                &&&&    SO 8(7)--7(6) &  12.6 $\pm$  2.4 & -16.90 &  13.9& \\
        L1     & $17^h45^m39.60^s$ & $-29^\circ01^\prime 04.29^{\prime \prime}$ & 3.67 &
              SiO 7--6 &  16.1 $\pm$  1.1 & -58.68 &  20.8&0.81 $\pm$ 0.07\\
        &&&&    SO 8(7)--7(6) &  29.1 $\pm$  1.6 & -58.91 &  14.2& \\
        L2     & $17^h45^m39.74^s$ & $-29^\circ01^\prime 05.41^{\prime \prime}$ & 2.25 &
              SiO 7--6 &  17.2 $\pm$  2.0 & -35.33 &  16.8&0.58 $\pm$ 0.08\\
        &&&&    SO 8(7)--7(6) &  31.7 $\pm$  2.2 & -35.60 &  15.8& \\
    \enddata
\end{deluxetable}

\begin{deluxetable}{ccccccccc}[htbp!]
    \centering
    \tablecaption{}
    \tablehead{
        Source & $\alpha$ & $\delta$ & Extent & Emission & Flux density & v & $\Delta v$ & $\mathrm{I}_\mathrm{SiO}/\mathrm{I}_\mathrm{SO}$ \\
        & (J2000) & (J2000) & (beam) & Line & (mJy/beam) & (km s$^{-1}$) & (km s$^{-1}$)  & 
    }
    \startdata
        L3     & $17^h45^m39.79^s$ & $-29^\circ01^\prime 5.57^{\prime \prime}$ & 1.02 &
              SiO 7--6 &  7.95 $\pm$  3.3 & -62.66 &  15.6&0.37 $\pm$ 0.16\\
        &&&&    SO 8(7)--7(6) &  22.4 $\pm$  3.4 & -61.95 &  15.1& \\
        L4     & $17^h45^m38.80^s$ & $-29^\circ01^\prime 06.84^{\prime \prime}$ & 2.18 &
              SiO 7--6 &  10.3 $\pm$  1.2 & -48.40 &  23.1&0.47 $\pm$ 0.06\\
        &&&&    SO 8(7)--7(6) &  28.2 $\pm$  1.6 & -51.13 &  18.0& \\
        L5     & $17^h45^m39.66^s$ & $-29^\circ01^\prime 06.15^{\prime \prime}$ & 1.99 &
              SiO 7--6 &  16.7 $\pm$  2.1 & -49.29 &  17.2&0.53 $\pm$ 0.08\\
        &&&&    SO 8(7)--7(6) &  22.4 $\pm$  1.5 & -51.23 &  24.0& \\
        L6     & $17^h45^m39.70^s$ & $-29^\circ01^\prime 07.10^{\prime \prime}$ & 5.11 &
              SiO 7--6 &  13.0 $\pm$ 0.71 & -55.02 &  24.5&0.6 $\pm$ 0.04\\
        &&&&    SO 8(7)--7(6) &  26.6 $\pm$ 0.87 &  -54.88 &  20.0& \\
        L7     & $17^h45^m39.75^s$ & $-29^\circ01^\prime 07.79^{\prime \prime}$ & 3.20 &
              SiO 7--6 &  18.9 $\pm$  1.7 & -61.44 &  17.9&0.43 $\pm$ 0.04\\
        &&&&    SO 8(7)--7(6) &  51.2 $\pm$  1.9 & -62.67 &  15.5& \\
        L8     & $17^h45^m39.68^s$ & $-29^\circ01^\prime 07.93^{\prime \prime}$ & 6.58 &
              SiO 7--6 &  10.1 $\pm$ 0.99 & -63.48 &  17.4&0.4 $\pm$ 0.04\\
        &&&&    SO 8(7)--7(6) &  34.2 $\pm$  1.3 & -64.93 &  13.0& \\
        L9     & $17^h45^m39.81^s$ & $-29^\circ01^\prime 7.82^{\prime \prime}$  & 2.25 &
              SiO 7--6 &  6.18 $\pm$  1.6 & -57.91 &  17.5&0.25 $\pm$ 0.07\\
        &&&&    SO 8(7)--7(6) &  28.4 $\pm$  1.8 & -60.78 &  15.3& \\
        L10a   & $17^h45^m39.81^s$ & $-29^\circ01^\prime 8.68^{\prime \prime}$ & 2.20 &
              SiO 7--6 &  12.1 $\pm$  2.4 & -60.44 &  15.2&0.34 $\pm$ 0.07\\
        &&&&    SO 8(7)--7(6) &  28.9 $\pm$  1.9 & -59.95 &  18.6& \\
        L10b   & $17^h45^m39.78^s$ & $-29^\circ01^\prime 08.75^{\prime \prime}$ & 2.80 &
              SiO 7--6 &  14.6 $\pm$  1.6 & -64.80 &  21.1&0.46 $\pm$ 0.06\\
        &&&&    SO 8(7)--7(6) &  39.1 $\pm$  2.0 & -64.58 &  17.2& \\
        L11   & $17^h45^m30.80^s$ & $-29^\circ01^\prime 09.66^{\prime \prime}$ & 3.50 &
              SiO 7--6 &  8.47 $\pm$  3.8 & -64.30 &  7.15&0.2 $\pm$ 0.09\\
        &&&&    SO 8(7)--7(6) &  17.3 $\pm$  1.5 & -69.77 &  17.7& \\
        L12   & $17^h45^m39.88^s$ & $-29^\circ01^\prime 9.93^{\prime \prime}$ & 1.78 &
              SiO 7--6 &  7.32 $\pm$  2.3 & -34.75 &  19.8&0.59 $\pm$ 0.22\\
        &&&&    SO 8(7)--7(6) &  15.0 $\pm$  2.8 & -35.23 &  16.3& \\
        L13   & $17^h45^m39.88^s$ & $-29^\circ01^\prime 4.67^{\prime \prime}$ & 5.74 &
              SiO 7--6 &  7.62 $\pm$ 0.31 & -46.78 &  36.6&0.51 $\pm$ 0.02\\
        &&&&    SO 8(7)--7(6) &  18.1 $\pm$ 0.38 & -46.09 &  30.0& \\
        L14   & $17^h45^m39.91^s$ & $-29^\circ01^\prime 4.06^{\prime \prime}$  & 4.39 &
              SiO 7--6 &  10.1 $\pm$ 0.59 & -58.36 &  24.2&0.62 $\pm$ 0.04\\
        &&&&    SO 8(7)--7(6) &  15.5 $\pm$ 0.55 & -55.07 &  25.6& \\
        L15   & $17^h45^m39.96^s$ & $-29^\circ01^\prime 3.62^{\prime \prime}$ & 1.78 &
              SiO 7--6 &  18.2 $\pm$  2.3 & -59.76 &  14.5&0.48 $\pm$ 0.07\\
        &&&&    SO 8(7)--7(6) &  30.8 $\pm$  1.9 & -58.14 &  17.7& \\
        L16   & $17^h45^m40.00^s$ & $-29^\circ01^\prime 3.38^{\prime \prime}$ & 1.70 &
              SiO 7--6 &  10.3 $\pm$  1.4 & -52.46 &  22.9&0.5 $\pm$ 0.08\\
        &&&&    SO 8(7)--7(6) &  26.9 $\pm$  1.8 & -51.29 &  17.5& \\
        L17   & $17^h45^m40.05^s$ & $-29^\circ01^\prime 2.33^{\prime \prime}$ & 4.53 & 
              SiO 7--6 &  30.9 $\pm$ 0.99 & -65.43 &  18.2&0.74 $\pm$ 0.03\\
        &&&&    SO 8(7)--7(6) &  35.1 $\pm$ 0.83 & -64.85 &  21.6& \\
        L18   & $17^h45^m40.09^s$ & $-29^\circ01^\prime 2.60^{\prime \prime}$ & 2.97 & 
              SiO 7--6 &  14.1 $\pm$ 0.86 & -49.59 &  23.6&0.59 $\pm$ 0.04\\
        &&&&    SO 8(7)--7(6) &  25.9 $\pm$ 0.92 & -49.22 &  21.9& \\
        L19   & $17^h45^m40.12^s$ & $-29^\circ01^\prime 1.78^{\prime \prime}$ & 4.07 & 
              SiO 7--6 &  9.37 $\pm$  1.3 & -45.00 &  13.8& 0.47 $\pm$ 0.07\\
        &&&&           &  28.3 $\pm$ 0.92 & -69.00 &  20.2& 0.80 $\pm$ 0.03\\
        &&&&    SO 8(7)--7(6) &  18.0 $\pm$  1.2 & -46.16 &  15.2& \\
        &&&&                 &  35.8 $\pm$ 0.92 & -68.97 &  20.0& \\
    \enddata
\end{deluxetable}

\begin{deluxetable}{ccccccccc}[htbp!]
    \centering
    \tablecaption{}
    \tablehead{
        Source & $\alpha$ & $\delta$ & Extent & Emission & Flux density & v & $\Delta v$ & $\mathrm{I}_\mathrm{SiO}/\mathrm{I}_\mathrm{SO}$ \\
        & (J2000) & (J2000) & (beam) & Line & (mJy/beam) & (km s$^{-1}$) & (km s$^{-1}$)  & 
    }
    \startdata
        L20   & $17^h45^m40.14^s$ & $-29^\circ01^\prime 1.13^{\prime \prime}$ & 2.73 &
              SiO 7--6 &  24.5 $\pm$  1.3 & -48.49 &  20.2&0.65 $\pm$ 0.04\\
        &&&&    SO 8(7)--7(6) &  48.4 $\pm$  1.7 & -48.64 &  15.8& \\
        L21   & $17^h45^m40.15^s$ & $-29^\circ01^\prime 00.44^{\prime \prime}$ & 3.22 &
              SiO 7--6 &  16.9 $\pm$  1.2 & -51.05 &  17.8&0.47 $\pm$ 0.04\\
        &&&&    SO 8(7)--7(6) &  31.4 $\pm$  1.1 & -52.57 &  20.4& \\
        L22   & $17^h45^m40.181^s$ & $-29^\circ01^\prime 00.89^{\prime \prime}$ & 2.52 &  
              SiO 7--6 &  26.4 $\pm$  1.4 & -47.16 &  19.5&0.54 $\pm$ 0.03\\
        &&&&    SO 8(7)--7(6) &  44.5 $\pm$  1.3 & -47.60 &  21.3& \\
        L23   & $17^h45^m40.19^s$ & $-29^\circ01^\prime 1.47^{\prime \prime}$ & 2.88 &
              SiO 7--6 &  15.2 $\pm$ 0.94 & -48.15 &  29.4&0.52 $\pm$ 0.04\\
        &&&&    SO 8(7)--7(6) &  32.6 $\pm$  1.1 & -48.62 &  26.2& \\
        L24   & $17^h45^m40.25^s$ & $-29^\circ01^\prime 00.94^{\prime \prime}$ & 4.02 &
              SiO 7--6 &  17.1 $\pm$  1.1 & -62.84 &  22.1&0.46 $\pm$ 0.03\\
        &&&&    SO 8(7)--7(6) &  37.1 $\pm$  1.1 & -62.44 &  22.0& \\
        L25   & $17^h45^m40.20^s$ & $-29^\circ01^\prime 00.14^{\prime \prime}$ & 2.94 &
              SiO 7--6 &  17.5 $\pm$ 0.92 & -61.57 &  29.6&0.66 $\pm$ 0.04\\
        &&&&    SO 8(7)--7(6) &  28.7 $\pm$  1.0 & -58.58 &  27.2& \\
        L26   & $17^h45^m40.27^s$ & $-29^\circ00^\prime 59.59^{\prime \prime}$ & 3.35 &
              SiO 7--6 &  12.4 $\pm$  1.2 & -52.84 &  22.7&0.53 $\pm$ 0.06\\
        &&&&    SO 8(7)--7(6) &  22.6 $\pm$  1.2 & -51.51 &  23.3& \\
        L27   & $17^h45^m40.28^s$ & $-29^\circ00^\prime 58.81^{\prime \prime}$ & 1.69 &
              SiO 7--6 &  17.7 $\pm$  2.8 & -30.47 &  18.8&0.56 $\pm$ 0.1\\
        &&&&    SO 8(7)--7(6) &  26.5 $\pm$  2.3 & -29.15 &  22.6& \\
        L28   & $17^h45^m40.46^s$ & $-29^\circ00^\prime 59.67^{\prime \prime}$ & 89.40 &
              SiO 7--6 &  12.7 $\pm$ 0.84 & -60.86 &  19.5&0.62 $\pm$ 0.05\\
        &&&&    SO 8(7)--7(6) &  18.7 $\pm$ 0.77 & -59.57 &  21.2& \\
        L29   & $17^h45^m40.19^s$ & $-29^\circ01^\prime 3.83^{\prime \prime}$ & 3.09 &
              SiO 7--6 &  7.54 $\pm$ 0.99 & -49.50 &  21.6&0.55 $\pm$ 0.08\\
        &&&&    SO 8(7)--7(6) &  12.2 $\pm$ 0.88 & -53.33 &  24.2& \\
        L30   & $17^h45^m40.47^s$ & $-29^\circ01^\prime 8.10^{\prime \prime}$ & 3.52 & 
              SiO 7--6 &  14.0 $\pm$  3.1 & -26.48 &  14.9& 0.89 $\pm$ 0.21\\
        &&&&           &  6.18 $\pm$  4.2 & -52.43 &  10.9& 13.46 $\pm$ 9.69\\
        &&&&    SO 8(7)--7(6) &  21.6 $\pm$  2.0 & -30.01 &  10.9& \\
        &&&&     &  12.7 $\pm$ 2.7 & -63.94 & 0.394& \\
        M1     & $17^h45^m41.31^s$ & $-29^\circ00^\prime 58.84^{\prime \prime}$ & 4.50 &
              SiO 7--6 &  8.67 $\pm$  1.0 & -45.25 &  14.4&0.65 $\pm$ 0.09\\
        &&&&    SO 8(7)--7(6) &  13.8 $\pm$  1.1 & -44.42 &  14.0& \\
        M2     & $17^h45^m41.41^s$ & $-29^\circ00^\prime 59.93^{\prime \prime}$ & 1.00 &
              SiO 7--6 &  2.98 $\pm$ 0.63 & -50.17 &  18.2&0.38 $\pm$ 0.09\\
        &&&&    SO 8(7)--7(6) &  8.22 $\pm$ 0.66 & -54.28 &  17.2& \\
        M3     & $17^h45^m41.38^s$ & $-29^\circ00^\prime 58.16^{\prime \prime}$ & 2.02 &
              SiO 7--6 &  6.39 $\pm$  1.9 & -42.64 &  15.4&0.43 $\pm$ 0.14\\
        &&&&    SO 8(7)--7(6) &  9.23 $\pm$  1.2 & -44.63 &  24.8& \\
        M4     & $17^h45^m41.23^s$ & $-29^\circ00^\prime 57.77^{\prime \prime}$ & 2.27 &
              SO 8(7)--7(6) &  10.8 $\pm$  3.2 & 23.10 &  8.21& \\
        &&&&    CH$_3$OH 2(1,1)--2(0,2) &  8.19 $\pm$  2.2 & 21.39 &  12.0& \\ 
    \enddata
\end{deluxetable}

\clearpage

\section{Inferred conditions} \label{sec:mle-conditions}

\begin{deluxetable}{ccccc}[htbp!]
    \centering
    \tablecaption{The points of highest posterior density (HPD) of the physical conditions towards each observational source. Sources with parameter entries marked with -- represent an unconstrained value in that dimension. \label{tab:mle-conditions}}
    \tablehead{
		Source & $\mathrm{T_{kin}}$ & $\log{\mathrm{n_{H}}}$ & $\log{\mathrm{N_{SiO}}}$ & $\log{\mathrm{N_{SO}}}$ \\ 
		& [\si{\kelvin}] & [\si{\per\centi\meter\cubed}] & [\si{\per\centi\meter\squared}] & [\si{\per\centi\meter\squared}] 
	}
		\startdata
		N8 & -- & -- & $13.30^{+0.85}_{-0.68}$ & $14.65^{+0.56}_{-0.23}$ \\
		N7a & -- & $6.73^{+0.26}_{-1.70}$ & $12.92^{+0.61}_{-0.13}$ & $14.202^{+0.421}_{-0.071}$ \\
		N7b & -- & $5.70^{+0.80}_{-0.72}$ & $12.830^{+0.378}_{-0.073}$ & $14.027^{+0.258}_{-0.049}$ \\
		C1 & $440^{+300}_{-240}$ & $5.74^{+0.62}_{-0.31}$ & $13.416^{+0.142}_{-0.053}$ & $14.593^{+0.144}_{-0.051}$ \\
		G1 & $310^{+490}_{-110}$ & $5.69^{+0.65}_{-0.51}$ & $13.45^{+0.22}_{-0.12}$ & $14.489^{+0.161}_{-0.100}$ \\
		G2 & $610^{+140}_{-180}$ & $5.88^{+0.16}_{-0.18}$ & $13.392^{+0.058}_{-0.064}$ & $14.500^{+0.075}_{-0.058}$ \\
		G3a & $81^{+159}_{-18}$ & $6.56\pm 0.29$ & $13.648^{+0.106}_{-0.095}$ & $14.717^{+0.115}_{-0.080}$ \\
		G3b & $108^{+182}_{-45}$ & $6.42^{+0.45}_{-0.27}$ & $13.710^{+0.114}_{-0.087}$ & $14.905^{+0.126}_{-0.062}$ \\
		G4 & -- & $6.21^{+0.56}_{-0.31}$ & $13.50^{+0.20}_{-0.13}$ & $14.639^{+0.199}_{-0.081}$ \\
		G5 & $63^{+22}_{-0}$ & $6.99^{+0.00}_{-0.31}$ & $13.981\pm 0.090$ & $14.985^{+0.055}_{-0.058}$ \\
		D1 & -- & -- & $13.59^{+0.40}_{-0.12}$ & $14.725^{+0.275}_{-0.069}$ \\
		D2 & $300^{+450}_{-100}$ & $5.76^{+0.49}_{-0.44}$ & $13.53^{+0.19}_{-0.11}$ & $14.704^{+0.115}_{-0.090}$ \\
		D3c1 & -- & $5.99^{+0.61}_{-0.75}$ & $13.13^{+0.31}_{-0.13}$ & $14.37^{+0.22}_{-0.10}$ \\
		D3c2 & $520^{+320}_{-310}$ & $5.81^{+0.57}_{-0.60}$ & $13.11^{+0.24}_{-0.14}$ & $14.417^{+0.178}_{-0.096}$ \\
		D4 & $297^{+509}_{-98}$ & $5.72^{+0.56}_{-0.52}$ & $13.408^{+0.199}_{-0.083}$ & $14.594^{+0.131}_{-0.057}$ \\
		D5 & $340^{+410}_{-210}$ & $5.78^{+0.67}_{-0.63}$ & $13.10^{+0.31}_{-0.15}$ & $14.55^{+0.19}_{-0.10}$ \\
		D6 & -- & $5.83^{+1.16}_{-0.76}$ & $13.33^{+0.56}_{-0.20}$ & $14.45^{+0.39}_{-0.16}$ \\
		D7 & -- & $6.40^{+0.59}_{-1.26}$ & $13.32^{+0.55}_{-0.21}$ & $14.47^{+0.37}_{-0.12}$ \\
		H1a & -- & $5.74^{+0.69}_{-0.50}$ & $13.558^{+0.225}_{-0.076}$ & $14.635^{+0.152}_{-0.089}$ \\
		H1b & -- & $5.90^{+0.65}_{-0.51}$ & $13.809^{+0.180}_{-0.069}$ & $14.713^{+0.154}_{-0.094}$ \\
		H2 & -- & $5.95^{+0.90}_{-1.34}$ & $13.57^{+0.74}_{-0.13}$ & $14.31^{+0.49}_{-0.20}$ \\
		H3 & -- & $6.84^{+0.15}_{-1.83}$ & $13.75^{+0.62}_{-0.13}$ & $14.52^{+0.43}_{-0.23}$ \\
		H4 & -- & $6.63^{+0.36}_{-1.48}$ & $13.774^{+0.556}_{-0.095}$ & $14.47^{+0.37}_{-0.12}$ \\
		J1 & -- & $5.67^{+1.05}_{-0.27}$ & $13.36^{+0.30}_{-0.17}$ & $14.38^{+0.21}_{-0.16}$ \\
		J2 & -- & -- & $14.239^{+0.316}_{-0.052}$ & $14.859^{+0.276}_{-0.051}$ \\
		J3 & -- & $5.69^{+0.85}_{-0.14}$ & $13.448^{+0.187}_{-0.088}$ & $14.593^{+0.175}_{-0.046}$ \\
		J4 & -- & -- & $13.29^{+0.45}_{-0.22}$ & $14.33^{+0.29}_{-0.18}$ \\
		K1 & $480^{+380}_{-240}$ & $5.93^{+0.68}_{-0.56}$ & $13.46^{+0.23}_{-0.12}$ & $14.61^{+0.19}_{-0.10}$ \\
		K2 & $450^{+370}_{-260}$ & -- & $13.39^{+0.42}_{-0.13}$ & $14.532^{+0.280}_{-0.099}$ \\
		K3 & -- & $5.67^{+1.32}_{-0.43}$ & $13.37^{+0.48}_{-0.16}$ & $14.614^{+0.321}_{-0.095}$ \\
		K4 & $460^{+290}_{-310}$ & $5.84^{+0.46}_{-0.68}$ & $13.45^{+0.24}_{-0.11}$ & $14.63^{+0.15}_{-0.10}$ \\
		K5a & -- & $5.61^{+1.16}_{-0.29}$ & $13.20^{+0.34}_{-0.16}$ & $14.57^{+0.25}_{-0.11}$ \\
		K5b & -- & $5.87^{+1.12}_{-0.56}$ & $13.24^{+0.41}_{-0.15}$ & $14.446^{+0.303}_{-0.081}$ \\
		K5c & -- & $5.87^{+0.85}_{-0.60}$ & $13.25^{+0.34}_{-0.17}$ & $14.47^{+0.24}_{-0.12}$ \\
		K6a & $360^{+400}_{-200}$ & $5.73^{+0.76}_{-0.47}$ & $13.51^{+0.27}_{-0.12}$ & $14.72^{+0.18}_{-0.10}$ \\
		K6b & -- & $5.82^{+0.99}_{-0.79}$ & $13.46^{+0.47}_{-0.14}$ & $14.668^{+0.294}_{-0.089}$ \\
		K7a & -- & $5.65^{+0.85}_{-0.32}$ & $13.594^{+0.188}_{-0.072}$ & $14.784^{+0.151}_{-0.042}$ \\
		K7b & -- & $5.9\pm 1.0$ & $13.44^{+0.61}_{-0.24}$ & $14.58^{+0.44}_{-0.20}$ \\
		K8a & $590^{+210}_{-340}$ & $5.78^{+0.46}_{-0.45}$ & $13.97^{+0.17}_{-0.11}$ & $15.108^{+0.127}_{-0.097}$ \\
		K8bc1 & -- & $6.20^{+0.79}_{-1.20}$ & $13.848^{+0.595}_{-0.097}$ & $14.58^{+0.39}_{-0.15}$ \\
		K8bc2 & $--$ & $5.84^{+1.15}_{-0.46}$ & $13.52^{+0.34}_{-0.16}$ & $14.703^{+0.254}_{-0.086}$ \\
	\enddata
\end{deluxetable} 

\begin{deluxetable}{ccccc}[htbp!]
    \tablehead{
		Source & $\mathrm{T_{kin}}$ & $\log{\mathrm{n_{H}}}$ & $\log{\mathrm{N_{SiO}}}$ & $\log{\mathrm{N_{SO}}}$ \\ 
		& [\si{\kelvin}] & [\si{\per\centi\meter\cubed}] & [\si{\per\centi\meter\squared}] & [\si{\per\centi\meter\squared}] 
	}
	    \startdata
	    K9 & -- & $6.09^{+0.53}_{-0.77}$ & $13.65^{+0.30}_{-0.12}$ & $14.907^{+0.215}_{-0.097}$ \\
		K10 & -- & $5.93^{+1.06}_{-0.63}$ & $13.58^{+0.44}_{-0.13}$ & $14.654^{+0.301}_{-0.088}$ \\
		K11 & -- & $5.77^{+1.22}_{-0.77}$ & $13.61^{+0.62}_{-0.32}$ & $14.83^{+0.38}_{-0.20}$ \\
		K12 & -- & -- & $13.49^{+0.73}_{-0.47}$ & $14.81^{+0.42}_{-0.21}$ \\
		K13a & -- & $6.01^{+0.82}_{-0.86}$ & $13.53^{+0.36}_{-0.18}$ & $14.787^{+0.263}_{-0.072}$ \\
		K13b & -- & -- & $13.68^{+0.73}_{-0.51}$ & $14.95^{+0.43}_{-0.22}$ \\
		K14 & $580^{+360}_{-290}$ & -- & $13.48^{+0.68}_{-0.31}$ & $14.59^{+0.50}_{-0.24}$ \\
		K15a & $470^{+400}_{-240}$ & $5.85^{+1.14}_{-0.65}$ & $13.43^{+0.44}_{-0.14}$ & $14.41^{+0.31}_{-0.14}$ \\
		K15b & $720^{+280}_{-360}$ & $6.16^{+0.83}_{-1.05}$ & $13.50^{+0.51}_{-0.21}$ & $14.63^{+0.35}_{-0.15}$ \\
		K16 & -- & $5.66^{+1.33}_{-0.60}$ & $13.47^{+0.57}_{-0.30}$ & $14.73^{+0.36}_{-0.14}$ \\
		K17 & -- & -- & $13.87^{+0.79}_{-0.31}$ & $14.84^{+0.63}_{-0.35}$ \\
		K18 & -- & $6.32^{+0.68}_{-1.11}$ & $13.11^{+0.48}_{-0.19}$ & $14.30^{+0.33}_{-0.10}$ \\
		K19 & -- & $5.81^{+1.17}_{-0.55}$ & $13.47^{+0.41}_{-0.11}$ & $14.494^{+0.286}_{-0.084}$ \\
		K20 & -- & $5.43^{+0.98}_{-0.45}$ & $13.46^{+0.37}_{-0.14}$ & $14.71^{+0.23}_{-0.12}$ \\
		K21a & -- & $5.82^{+1.17}_{-0.47}$ & $13.82^{+0.38}_{-0.13}$ & $14.931^{+0.263}_{-0.086}$ \\
		K21b & $610^{+270}_{-380}$ & $6.04^{+0.92}_{-1.08}$ & $13.54^{+0.59}_{-0.22}$ & $14.70^{+0.39}_{-0.17}$ \\
		K22 & -- & -- & $13.50^{+0.86}_{-0.58}$ & $14.65^{+0.63}_{-0.33}$ \\
		K23 & -- & $5.88^{+1.07}_{-0.66}$ & $13.01^{+0.51}_{-0.25}$ & $14.31^{+0.35}_{-0.13}$ \\
		K24 & $917^{+81}_{-574}$ & $5.74^{+1.25}_{-0.93}$ & $13.31^{+0.75}_{-0.25}$ & $14.40^{+0.53}_{-0.24}$ \\
		L1 & $610^{+190}_{-420}$ & $5.65^{+0.67}_{-0.45}$ & $13.66^{+0.25}_{-0.11}$ & $14.739^{+0.160}_{-0.096}$ \\
		L2 & -- & $5.75^{+0.63}_{-0.52}$ & $13.61^{+0.25}_{-0.14}$ & $14.84^{+0.16}_{-0.12}$ \\
		L3 & $480^{+360}_{-320}$ & $5.85^{+1.00}_{-0.91}$ & $13.29^{+0.67}_{-0.52}$ & $14.68^{+0.41}_{-0.17}$ \\
		L4 & $580^{+230}_{-410}$ & $5.68^{+0.80}_{-0.47}$ & $13.52^{+0.26}_{-0.13}$ & $14.828^{+0.189}_{-0.095}$ \\
		L5 & -- & -- & $13.60^{+0.37}_{-0.12}$ & $14.843^{+0.271}_{-0.063}$ \\
		L6 & $560^{+220}_{-330}$ & $5.77^{+0.58}_{-0.31}$ & $13.631^{+0.156}_{-0.093}$ & $14.85^{+0.13}_{-0.10}$ \\
		L7 & -- & $5.69^{+0.63}_{-0.53}$ & $13.68^{+0.24}_{-0.13}$ & $15.029^{+0.160}_{-0.095}$ \\
		L8 & -- & $5.76^{+0.66}_{-0.42}$ & $13.38^{+0.20}_{-0.12}$ & $14.78^{+0.15}_{-0.11}$ \\
		L9 & -- & $5.63^{+1.08}_{-0.42}$ & $13.20^{+0.38}_{-0.24}$ & $14.78^{+0.25}_{-0.12}$ \\
		L10a & $580^{+310}_{-340}$ & $5.46^{+1.07}_{-0.41}$ & $13.41^{+0.35}_{-0.18}$ & $14.86^{+0.24}_{-0.10}$ \\
		L10b & -- & $5.78^{+0.58}_{-0.59}$ & $13.64^{+0.25}_{-0.13}$ & $14.96^{+0.16}_{-0.10}$ \\
		L11 & -- & -- & $12.97^{+0.54}_{-0.51}$ & $14.604^{+0.324}_{-0.076}$ \\
		L12 & -- & $6.20^{+0.79}_{-1.26}$ & $13.35^{+0.64}_{-0.35}$ & $14.55^{+0.44}_{-0.21}$ \\
		L13 & $510^{+260}_{-270}$ & $5.72^{+0.50}_{-0.37}$ & $13.58^{+0.17}_{-0.10}$ & $14.847^{+0.112}_{-0.085}$ \\
		L14 & $420^{+340}_{-240}$ & $5.79^{+0.65}_{-0.42}$ & $13.513^{+0.200}_{-0.090}$ & $14.721^{+0.148}_{-0.095}$ \\
		L15 & $350^{+410}_{-190}$ & $5.71^{+0.78}_{-0.52}$ & $13.56^{+0.27}_{-0.14}$ & $14.863^{+0.195}_{-0.096}$ \\
		L16 & $480^{+390}_{-240}$ & $5.79^{+0.93}_{-0.39}$ & $13.51^{+0.28}_{-0.13}$ & $14.81^{+0.21}_{-0.11}$ \\
		L17 & $500^{+300}_{-230}$ & $5.78^{+0.50}_{-0.33}$ & $13.876^{+0.135}_{-0.075}$ & $14.998^{+0.113}_{-0.083}$ \\
		L18 & -- & $5.83^{+0.52}_{-0.46}$ & $13.66^{+0.17}_{-0.10}$ & $14.869^{+0.144}_{-0.092}$ \\
		L19c1 & $680^{+320}_{-290}$ & $5.81^{+0.62}_{-0.77}$ & $13.26^{+0.31}_{-0.16}$ & $14.559^{+0.230}_{-0.097}$ \\
		L19c2 & -- & $5.88^{+0.47}_{-0.41}$ & $13.886^{+0.146}_{-0.082}$ & $14.979^{+0.118}_{-0.088}$ \\
		L20 & $500^{+290}_{-260}$ & $5.88^{+0.46}_{-0.44}$ & $13.828^{+0.147}_{-0.094}$ & $15.002^{+0.137}_{-0.083}$ \\
		L21 & $560^{+190}_{-410}$ & $5.80^{+0.55}_{-0.53}$ & $13.61^{+0.21}_{-0.10}$ & $14.928^{+0.145}_{-0.086}$ \\
		L22 & -- & $5.75^{+0.62}_{-0.32}$ & $13.845^{+0.172}_{-0.093}$ & $15.095^{+0.137}_{-0.083}$ \\
		L23 & -- & $5.77^{+0.61}_{-0.38}$ & $13.781^{+0.162}_{-0.090}$ & $15.052^{+0.134}_{-0.089}$ \\
		L24 & $590^{+140}_{-370}$ & $5.82^{+0.47}_{-0.43}$ & $13.706^{+0.180}_{-0.098}$ & $15.040^{+0.119}_{-0.099}$ \\
		\enddata
\end{deluxetable}

\begin{deluxetable}{ccccc}[htbp!]
    \tablehead{
		Source & $\mathrm{T_{kin}}$ & $\log{\mathrm{n_{H}}}$ & $\log{\mathrm{N_{SiO}}}$ & $\log{\mathrm{N_{SO}}}$ \\ 
		& [\si{\kelvin}] & [\si{\per\centi\meter\cubed}] & [\si{\per\centi\meter\squared}] & [\si{\per\centi\meter\squared}] 
	}
	    \startdata
		L25 & $300^{+420}_{-130}$ & $5.74^{+0.45}_{-0.51}$ & $13.85^{+0.23}_{-0.11}$ & $15.016^{+0.122}_{-0.094}$ \\
		L26 & $580^{+260}_{-350}$ & $5.86^{+0.67}_{-0.56}$ & $13.58^{+0.24}_{-0.11}$ & $14.85^{+0.18}_{-0.11}$ \\
		L27 & $360^{+430}_{-210}$ & $5.82^{+0.68}_{-0.57}$ & $13.67^{+0.28}_{-0.17}$ & $14.92^{+0.18}_{-0.13}$ \\
		L28 & $350^{+460}_{-150}$ & $5.89^{+0.55}_{-0.54}$ & $13.54^{+0.19}_{-0.11}$ & $14.719^{+0.152}_{-0.097}$ \\
		L29 & -- & $5.79^{+0.72}_{-0.55}$ & $13.35^{+0.26}_{-0.15}$ & $14.59^{+0.20}_{-0.11}$ \\
		L30c1 & $490^{+410}_{-240}$ & $5.99^{+0.89}_{-0.68}$ & $13.48^{+0.43}_{-0.21}$ & $14.50^{+0.30}_{-0.11}$ \\
		L30c2 & $450^{+410}_{-270}$ & -- & $13.02^{+0.62}_{-1.01}$ &  \\
		M1 & $400^{+460}_{-170}$ & $5.65^{+0.87}_{-0.58}$ & $13.23^{+0.36}_{-0.14}$ & $14.42^{+0.23}_{-0.12}$ \\
		M2 & $510^{+370}_{-290}$ & -- & $12.89^{+0.39}_{-0.19}$ & $14.27^{+0.28}_{-0.10}$ \\
		M3 & -- & $6.78^{+0.17}_{-1.49}$ & $13.17^{+0.51}_{-0.27}$ & $14.50^{+0.35}_{-0.13}$ \\
		\enddata
\end{deluxetable}

% \allauthors


\end{document}